\documentclass{aa}
\usepackage{amsmath}
\usepackage[varg]{txfonts}
\usepackage{amssymb}
\usepackage{mathrsfs}
\usepackage{graphicx}
\usepackage{hyperref}
\usepackage{natbib}
\bibpunct{(}{)}{;}{a}{}{,} 
\usepackage{afterpage,array,longtable}

\usepackage{microtype}
\usepackage{wasysym}
\usepackage{colortbl}
\usepackage{xcolor}

\def\parcmin{\hbox{$^\prime\phantom{^\prime}$}}
\def\magg{\ensuremath{\text{mag}\,\text{arcsec}^{-2}}}

\def\King{K71}
\def\CV{CV83}
\def\MBH{MBH94}
\def\Lequeux{LFD96}
\def\Zheng{ZSS99}
\def\Bernstein{B07}
\def\Jong{J08}
\def\Slater{SHM09}

\def\PSFB {PSF$_{\text{B07}}$}
\def\PSFS {PSF$_{\text{S09}}$}
\def\PSFAbr{PSF$_{\text{A14}}$}
\def\PSFC {PSF$_{\text{CV}}$}
\def\PSFCa{PSF$_{\text{CV,1}}$}
\def\PSFCb{PSF$_{\text{CV,2}}$}
\def\PSFK {PSF$_{\text{K71}}$}
\def\PSFKo{PSF$_{\text{K73}}$}
\def\PSFPa{PSF$_{\text{P,LR}}$}
\def\PSFPi{PSF$_{\text{P73}}$}
\def\PSFVa{PSF$_{V,\text{0m}}$}
\def\PSFVb{PSF$_{V,\text{3m}}$}
\def\PSFIa{PSF$_{i,\text{0m}}$}
\def\PSFIb{PSF$_{i,\text{3m}}$}
\def\PSFMBH{PSF$_{\text{MBH}}$}
\def\PSFMBHT{PSF$_{\text{MBH}}^{\text{new}}$}

\def\sd{\ensuremath{\phantom{0}}}
\def\sdd{\ensuremath{\phantom{00}}}

\begin{document}

\title{The influence of diffuse scattered light}
\subtitle{I. The PSF and its role to observations of the edge-on galaxy NGC 5907}
\authorrunning{C.\ Sandin}

\author{Christer~Sandin\thanks{E-mail: CSandin@aip.de}}
\institute{Leibniz-Institut f\"ur Astrophysik Potsdam (AIP), An der Sternwarte 16, 144 82 Potsdam, Germany}

\date{Received 15 January 2014 / Accepted 12 June 2014}

\abstract{
All telescopes and instruments are to some degree affected by scattered light. It is possible to estimate the amount of such scattered light, and even correct for it, with a radially extended point spread function (PSF). The outer parts of the PSF have only rarely been determined, since they are faint and therefore difficult to measure. A mostly complete overview of existing properties and measurements of radially extended PSFs is presented, to both show their similarities and to indicate how bright extended objects can be used to measure the faintest regions. The importance of the far wings of the PSF and their possible temporal variations are demonstrated in three edge-on galaxy models. The same study is applied to the first edge-on galaxy where earlier observations reveal a halo, NGC 5907. All PSFs were collected in two diagrams, after they were offset or normalized, when that was possible. Surface-brightness structures of edge-on galaxies were modelled and analysed to study scattered-light haloes that result with an exponential disc. The models were convolved with both a lower-limit PSF and a more average PSF. The PSF of the observed data could be used in the case of NGC 5907. The comparison of the PSFs demonstrates a lower-limit $r^{-2}$ power-law decline at larger radii. The analysis of the galaxy models shows that also the outer parts of the PSF are important to correctly model and analyse observations and, in particular, fainter regions. The reassessed analysis of the earlier measurements of NGC 5907 reveals an explanation for the faint halo in scattered light, within the quoted level of accuracy.
}


\maketitle


\section{Introduction}\label{sec:introduction}
Optical parts of telescopes, instruments, and detectors, as well as the atmosphere, give rise to scattered light. Imaging theory defines a point spread function (PSF) that describes how the light of a point source is affected by various optical scattering effects within these parts. The projected surface brightness structure of an object is convolved with the PSF to form the observed structure. The time-variable and field-dependent PSF, moreover, extends to large angular radii. Whilst the PSF rapidly becomes faint with increasing radii, the integrated amount of light in its faint extended wings can still be significant. If not corrected for, the scattered light adds a systematic component to observed intensities, where the amplitude, the shape, and its influence on data are unclear.

Various astronomical studies examine, to different depths of detail, how scattered light influences their analysis. The first studies address large elliptical galaxies \citep{Va:48,Va:53}, \object{M\,31} \citep{Va:58}, and \object{NGC\,3379} \citep[hereafter \CV]{VaCa:79,CaVa:83}; in these cases effects of scattered light are small. Corresponding effects on envelopes of supergiant elliptical (cD) galaxies appear to be small as well \citep{Mac:92}; however, the light profile in the outer regions of \object{A\,2029} closely follows the standard elliptical-galaxies de Vaucouleur's law \citep{UsBoKu:91}, but only after they carefully remove extended and diffuse scattered-light components of field stars.

To isolate effects on colours, it is necessary to consider possible variations of the PSF with the wavelength. One study of 36 elliptical and lenticular galaxies convolve data of one bandpass with the PSF of the second bandpass before calculating colours \citep{IdMiFr:02}; the study, however, provides no information on how this method compares with a deconvolution in the separate bandpasses. The combined effects of colours and temporal variations of scattered light are at first studied in observations of four elliptical-type galaxies \citep{Mi:02}; this important study reports smaller effects in the two larger galaxies \object{NGC\,4406} and \object{NGC\,4473}, whilst the two smaller galaxies \object{NGC\, 4550} and \object{NGC\,4551} show larger effects.

Observations of extended emission around much brighter point sources are easily corrected through the use of a scaled PSF. Several studies subtract substantial amounts of scattered light from the central-star component in observations of circumstellar shells around old giant stars \citep{BeLa:75,MaCa:92,PlLa:94,GuErKi.:97,DeOlSc.:01}. Observations of hosts of distant quasi-stellar objects are also corrected by subtracting emission that originates in one or a few bright components that appear as point sources \citep{RoGrWaOr:96,WiScBr.:02,OrRoGr:03}. Extended and faint ionized haloes around planetary nebulae (PNe) are affected by scattered light from the drastically brighter central parts \citep{MiClWa:89}; the authors do not measure a PSF by themselves, but analyse their data of \object{BD$+30\degr3639$} with the extended PSF of \citet[hereafter \King]{Ki:71}. Other studies search deep exposures of PNe for additional haloes \citep[e.g.,][]{CoScStPe:03}, focusing on observations that show clumps and asymmetries, and avoiding data that only show diffuse light. With few exceptions these studies do not derive any physical properties of the haloes. One bold spectroscopic study, so far, both addresses scattered light and aims to measure physical properties in PN haloes \citep{SaScRo.:08}; the goal of this study was to derive the mass-loss evolution of the previous phase on the asymptotic giant branch with the PN halo data.

De Jong (\citeyear{Jo:08}, hereafter \Jong) makes a more detailed study of the influence of scattered light on observations of smaller edge-on galaxies, both with data in the \textit{Hubble} Ultra Deep Field \citep{BeStKo.:06} of the \textit{Hubble} Space Telescope (\textit{HST}) and in stacked data of the ground-based Sloan Digital Sky Survey (SDSS). He argues that PSF effects nearly fully explain \textit{HST} data, but effects are smaller in SDSS data (accounting for 20--80 per cent of the halo light) and merely inner regions are affected. He dismisses its role in larger objects by their size, because he assumes a too steep slope at large radii (see below). \citet{TzScLuSo:09} set out to study density profiles in the halo outside a shock front around the supernova remnant of the Crab nebula, but because of high levels of scattered light, this was unattainable. \citet{BeZaCa:10} make a case to dismiss scattered light as a general phenomenon in a simplified study of stacked SDSS images of low surface-brightness edge-on galaxies. The most recent study I mention is that of \citet{FeHaCi.:13}, who make a serious effort to account for effects of scattered light in a search for haloes around \mbox{Ly$\alpha$} emitting galaxies; they find that haloes at redshift $z\simeq2.1$ can be fully explained by scattered light, whilst small haloes of galaxies at redshift $z\simeq3.1$ are partly still present after the scattered light is removed.

The cases that are mentioned above cannot be more than examples, as how scattered light is treated is nearly always part of the methods, and not a main topic. With few exceptions, studies that claim they address the role of scattered light only focus on measuring one PSF for all data, which is scaled to see how it matches object intensity structures of largely different origins. This procedure is ineffective with extended objects, where it is necessary to deconvolve surface-brightness structures with the PSF at the time of the observations. The few studies that deconvolve their data rarely consider temporal variations in the outer parts of the PSF, or they underestimate or truncate the PSF at some radius. The extended PSF is central to the analysis of scattered light, but it is unclear how it varies with wavelength, time, and location. And the faint outer wings are poorly known, partly since there is no established procedure to measure them.

This study focuses on ground-based observations in the visual wavelength range, 300--900nm. I begin with an overview of measured radially extended PSFs in Sect.~\ref{sec:PSF}, providing an update of the studies of {\King} and \citet[hereafter \Bernstein]{Be:07}. The overview focuses on two aspects that were not accounted for in any detail before. Measurements of the encircled energy show that the far regions of the PSF may contribute significant amounts of light. Current measurements of PSFs only measure stars; I demonstrate how planets, the Moon, and the Sun, can be used at larger radii instead, as in earlier studies. The analysis method that is used to model surface-brightness structures of edge-on galaxies is described in Sect.~\ref{sec:methods}. Example models of a small, an intermediate size, and a large edge-on disc galaxy are also analysed here; these examples demonstrate the decisive importance of using PSFs that are not truncated at short radii.

As an example of a real object, I analyse models and measurements of the edge-on galaxy \object{NGC 5907} in Sect.~\ref{sec:ngc5907}. This is an important object, since it presents the first case where a halo of excess light was found around an edge-on galaxy \citep{SaMoHaBo:94}. I reassessed the analysis of \citet[hereafter \MBH]{MoBoHa:94}, to show that it is possible to explain both the halo and the red excess in the halo by scattered light. The paper is finished with a brief discussion and conclusions in Sect.~\ref{sec:conclusions}.

\section{Measurements and properties of the PSF}\label{sec:PSF}
In the formalism of imaging theory the PSF describes how a point-source image is affected by broadening through detector effects, optical aberrations, diffraction, and scattering effects within the instrument, the telescope, and the atmosphere. The shape of the PSF is determined by atmospheric turbulence in the bright centre region, where it can be described by a Moffat profile, using Kolmogorov statistics \citep{Ra:96}. The much fainter region outside the centre, the so-called aureole, is less well understood and has been a target for scrutiny. Possible sources of the aureole include scattering by atmospheric aerosols and dust, as well as micro-ripples and dust on optical surfaces \citep[for example,][]{Hu:48,De:57,De:59}, and effects of diffusion and reflection within the instrument (\citealt{HaBu:95}; \citealt{Ra:96}; \Bernstein; \citealt{SlHaMi:09}, hereafter \Slater). The aureole eventually, at say $1\fdg5$, merges into an extremely faint `blue-sky' component of Rayleigh scattering that extends to, say, $90^{\circ}$ \citep{De:57,De:59}.

I begin with an overview and comparison of empirically measured radially extended PSFs in Sect.~\ref{sec:lpsf}. The PSF comparison of {\Bernstein} is discussed separately in Sect.~\ref{sec:Bernstein}. Thereafter, I describe how the PSFs are normalized and extrapolated to larger radii in Sect.~\ref{sec:enpsf}, and discuss the outcome of measurements of the encircled energy in Sect.~\ref{sec:eepsf}. Guidelines on how to make new PSF measurements are provided in Sect.~\ref{sec:newpsf}. Only instrument and telescope effects remain outside of the Earth atmosphere. A brief outline to how \textit{HST} data could also be affected by radially extended PSFs is presented for completeness in Sect.~\ref{sec:HSTpsf}.

\begin{table*}
\caption{Chronological list of measurements of radially extended PSFs.}
\centering
\label{tab:psf}
\tabcolsep=4.1pt
\begin{tabular}{lllllllll}\hline\hline\\[-1.8ex]
PSF & Fig. & Ref. & Observatory & Alt. & Telescope & Band / Line & Range & Object (Symbol)\\[1.0ex]\hline\\[-1.8ex]
\PSFK & \ref{fig:psf}, \ref{fig:bpsf} & 1--4 & Solar Physics & {\sdd}30 & 36" Common reflector & \ldots  & $\sim20\arcsec$--$180\arcsec$  & stars\\
      & \ref{fig:psf} & 2--4  &               &    &                      & \ldots  & 1\arcmin, 4\arcmin & $\astrosun$ (\textcolor{magenta}{$\triangle$})\\
      & \ref{fig:psf}, \ref{fig:bpsf} & 3, 4  & Lowell & 2195 & 21" reflector & $B$ & $\sim20\arcsec$--$5^{\circ}$  & $\jupiter$ and 2m star\\
      & \ref{fig:psf}, \ref{fig:bpsf} & 4  & Palomar & 1713 & 48" Schmidt & $B$ & $1\farcs3$--$228\arcsec$  & stars (\textcolor{magenta}{$\bullet$}, \textcolor{magenta}{$\times$})\\
      & \ref{fig:psf}, \ref{fig:bpsf} &   & Mount Wilson & 1742  & 60" reflector & $B$ & --$5\farcs6$  & 3 stars (\textcolor{magenta}{$\circ$})\\
\rowcolor[gray]{0.9}
\PSFKo & \ref{fig:psf}, \ref{fig:bpsf} & 5 & Palomar & 1713 & 48" Schmidt & $B$ & $7\arcsec$--$1^{\circ}$ & stars\\
\PSFPi & \ref{fig:psf} & 6 & Goethe Link & {\sd}293 & 16" reflector & $B$ & $48\arcsec$--$170\arcmin$ & stars, $\leftmoon$ (\textcolor{blue}{$\circ$}, \textcolor{blue}{$\bullet$}, \textcolor{blue}{\small$\blacktriangle$})\\
       & &  & McDonald    & 2076 & 2.1m Otto Struve & \ldots & not shown & \\
\rowcolor[gray]{0.9}
PSF$_{\text{S74}}$ & \ref{fig:bpsf} & 7 
          & Palomar & 1713 & 48" Schmidt & RG-1 & 100--600\arcsec & stars (\textcolor{red}{\small$\blacktriangle$})\\
\PSFC, &               & 8
         & McDonald & 2076 & 0.9m & $B$ & $10\arcsec$--$4\farcm6$ & $\gamma$CMa\\
(\PSFCa, & \ref{fig:psf} &               &          &      &      & \ldots & $3.5$--$90\arcmin$ & $\alpha$CMa\\
$\phantom{(}$\PSFCb) & \ref{fig:psf}, \ref{fig:bpsf} &
         &          &      &      & \ldots & $1.5$--$90^{\circ}$ & $\astrosun$\\
\rowcolor[gray]{0.9}
PSF$_{\text{S90}}$ & \ref{fig:bpsf} & 9
          & Calar Alto     & 2168 & 1.23m                  & $R$ & $\la480\arcsec$ & stars ({\small$\Box$})\\
PSF$_{\text{U91}}$ & \ref{fig:bpsf} & 10
          & Kitt Peak      & 2160 & 1.09m & $R$ & 1--$975\arcsec$ & stars\\
\rowcolor[gray]{0.9}
PSF$_{\text{M92}}$ & \ref{fig:bpsf} & 11
          & Kitt Peak      & 2160 & 24/36" B.\ Schmidt & $g$ & $5$--$92\arcsec$ & HD 19445 (\textcolor{green}{$\bullet$})\\
\PSFMBH & \ref{fig:psf} & 12
          & Kitt Peak      & 2160 & No.~1 0.9m (SARA)  & Harris $R$ & 0--150\arcsec & two stars\\
\PSFMBHT & \ref{fig:psf} &
          &                &      &                    &            &               &          \\
\rowcolor[gray]{0.9}
\PSFVa, & \ref{fig:psf}, \ref{fig:bpsf} & 13
          & Haute Provence & {\sd}650 & 1.2m Newtonian & $V$ & $\la160\arcsec$ & stars\\
\rowcolor[gray]{0.9}
\PSFVb & \ref{fig:psf} & & & & & & &\\
\rowcolor[gray]{0.9}
\PSFIa, & \ref{fig:psf}, \ref{fig:bpsf} & 
          &                &     &                & $I$ & $\la160\arcsec$ & stars\\
\rowcolor[gray]{0.9}
\PSFIb & \ref{fig:psf} & & & & & & &\\
PSF$_{\text{G05}}$ & \ref{fig:bpsf} & 14
          & Las Campanas   & 2282 & 40" Swope  & Gunn $i$ & 10--400\arcsec & stars\\
\rowcolor[gray]{0.9}
{\PSFB} & \ref{fig:bpsf} & 15
          & Las Campanas   & 2282 & 100" Du Pont  & $r$ & $\la400\arcsec$ & stars\\
\PSFPa & \ref{fig:psf} & 16
         & Calar Alto & 2168 & 3.5m / PMAS & \mbox{H$\alpha$} & --$25\arcsec$ & $\alpha$Lyr\\
\rowcolor[gray]{0.9}
\PSFS & \ref{fig:psf} & 17
        & Kitt Peak & 2160 & 24/36" B.\ Schmidt                & Wash. $M$ & $1\farcs4$--$64\arcmin$ & $\alpha$Boo, star\\
SDSS  & \ref{fig:psf} & 18, 19,
        & Apache Point & 2788 & 2.5m & $g$, $r$, $i$, ... & $\la46\arcsec$ & stars\\
      && 20, 21 & & & & & & \\
\rowcolor[gray]{0.9}
\PSFAbr & \ref{fig:psf} & 22
        & New Mexico Skies & 2200 & Dragonfly Array & $r$ & 5\farcs1--56\farcm8 & $\alpha$Lyr\\
\hline
\end{tabular}
\tablefoot{Column.~1, PSF notation used in this paper; Col.~2, the PSF is shown in this figure; Col.~3, references; Col.~4, observatory; Col.~5, altitude of the observatory (m); Col.~6, telescope (/instrument); Col.~7, wavelength bandpass or line that was used; Col.~8; radial extent of PSF measurements; and Col.~9, objects used to measure the PSF (and symbols used in Fig.~\ref{fig:psf}). The object symbols are $\jupiter$ (Jupiter), $\leftmoon$ (the Moon), and $\astrosun$ (the Sun).}
\tablebib{(1) \citet{ReSh:38}; (2) \citet{Hu:48}; (3) \citet{Va:58}; (4) \King; (5) \citet{Ko:73}; (6) \citet{Pi:73}; (7) \citet{Sh:74}; (8) {\CV}; (9) \citet{SuSeBe:90}; (10) \citet{UsBoKu:91}, (11) \citet{Mac:92}; (12) {\MBH}; (13) \citet{Mi:02}; (14) \citet{GoZaZa:05}; (15) \Bernstein; (16) \citet{SaScRo.:08}; (17) \Slater; (18) \citet{ZiWhBr:04}; (19) \Jong; (20) \citet{BeZaCa:10}; (21) \citet{TaDo:11}; (22) \citet{AbDo:14}
.}
\end{table*}

\subsection{An overview of measured radially extended PSFs}\label{sec:lpsf}
\begin{figure*}
\centering
\includegraphics{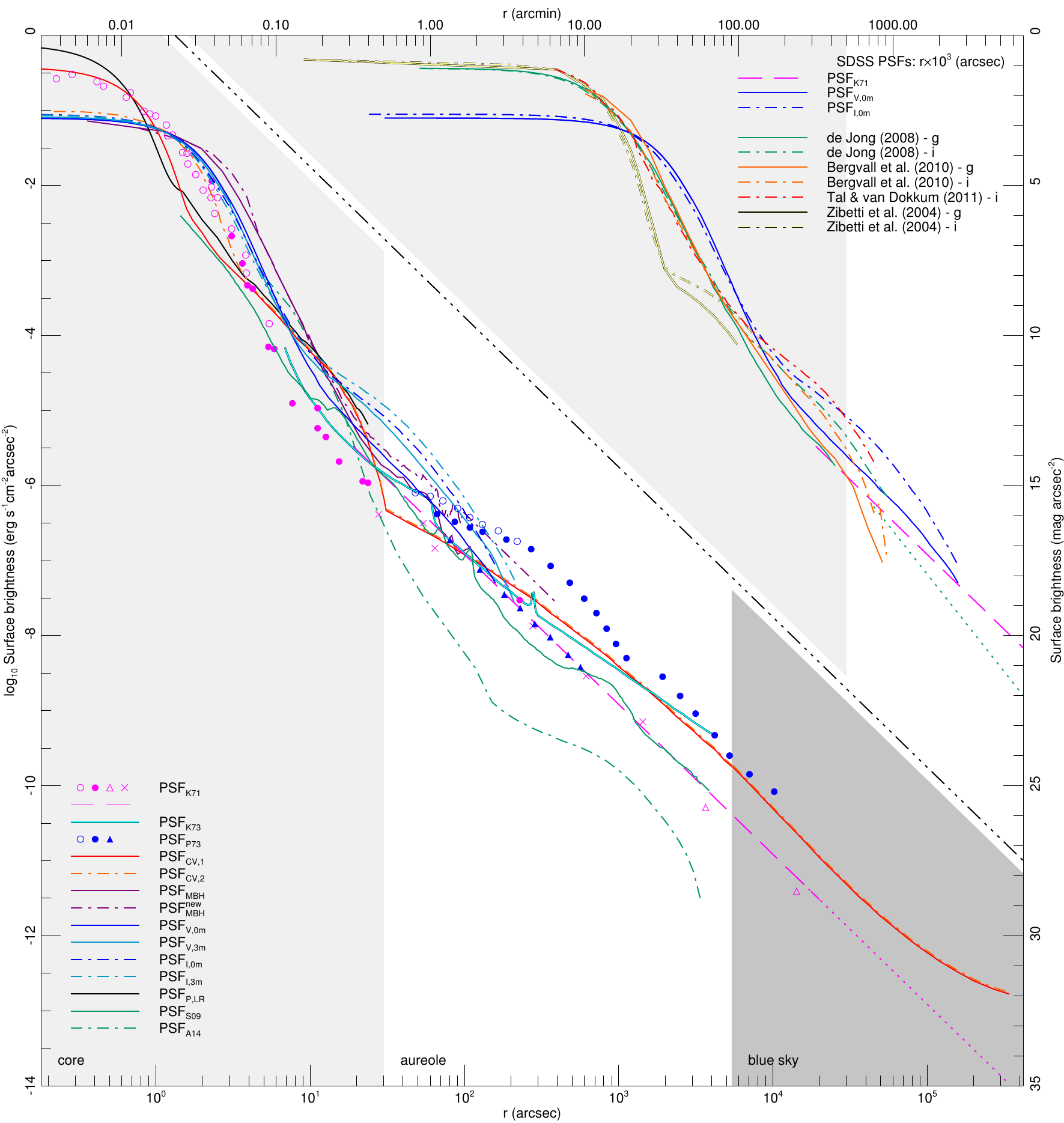}
\caption{PSF surface-brightness profiles versus the radius $r$ for a 0\,mag point source. Individual PSFs are drawn with coloured lines and symbols as indicated in the figure, also see Table~\ref{tab:psf}. PSFs of stacked SDSS images are shown in the upper part of the figure, as function of $10^{3}\times r$; {\PSFVa}, {\PSFIa}, and the outer regions of {\PSFK} are shown in both parts as references. Extrapolated PSFs are shown with dotted lines. The light (medium) grey region indicates the PSF core (blue sky) and the white region the aureole, as defined for {\PSFCa} and {\PSFCb} (\CV).}
\label{fig:psf}
\end{figure*}

\begin{figure*}
\sidecaption
\includegraphics{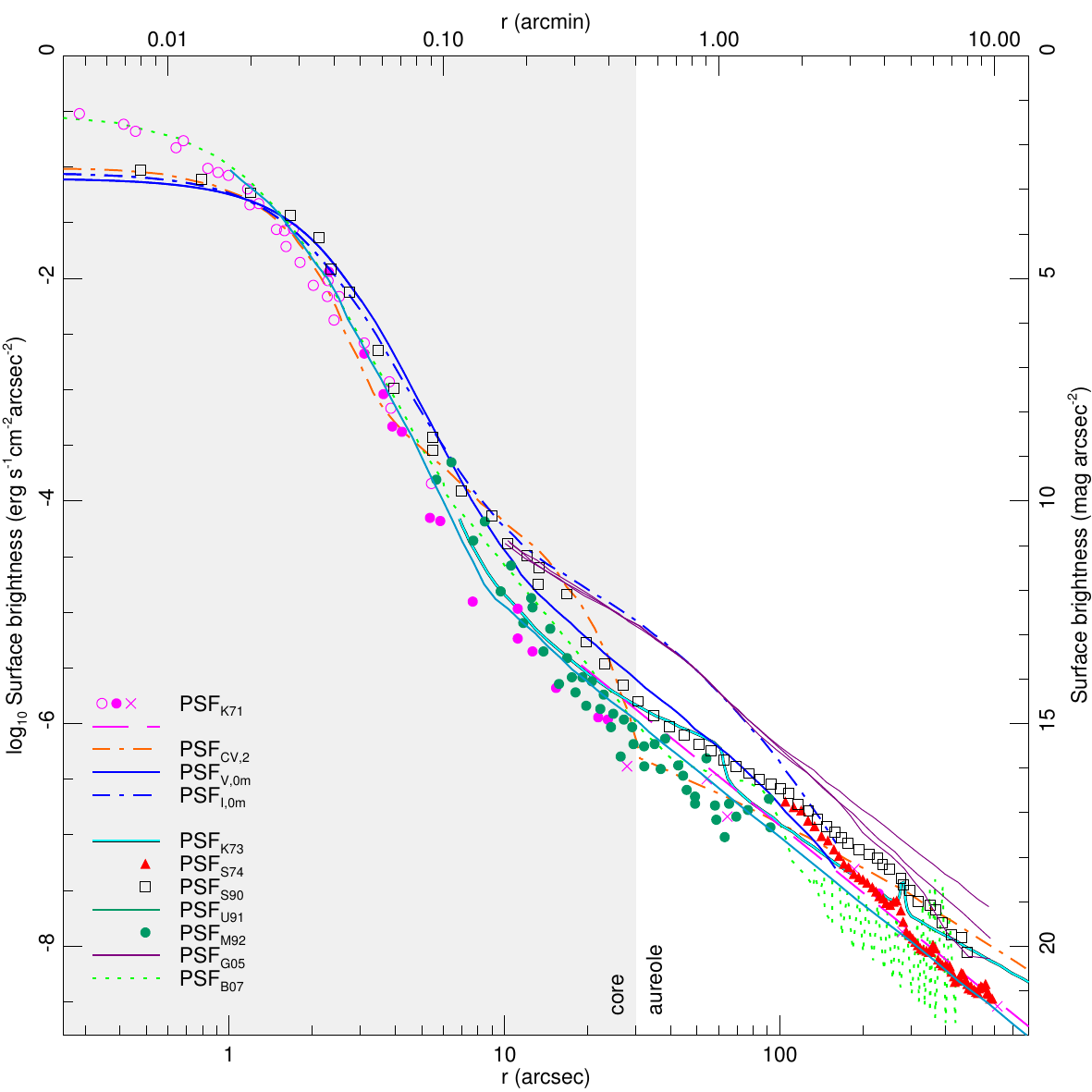}
\caption{PSF surface-brightness profiles versus the radius $r$ for a 0\,mag point source, from references that are discussed by {\Bernstein}. Individual PSFs are drawn with coloured lines and symbols as indicated in the figure, also see Table~\ref{tab:psf}. {\PSFK}, {\PSFVa}, {\PSFIa}, and {\PSFCb} are shown in this figure as references. The light grey region indicates the PSF core, and the white region the aureole, as defined for {\PSFCa} and {\PSFCb} (\CV).}
\label{fig:bpsf}
\end{figure*}

Few PSFs are measured out to large angles (radii $r$). Here I overview a sample of PSFs that were measured with different instruments and filters, at various telescopes, to show their similarities in terms of shape and radial extent. Most PSFs are shown in Fig.~\ref{fig:psf}, where they are normalized to 0\,mag; the PSFs that are discussed in Sect.~\ref{sec:Bernstein} are shown separately in Fig.~\ref{fig:bpsf}, to avoid that Fig.~\ref{fig:psf} is overfilled. Information of observational setups and radial extents of all discussed PSFs are collected in Table~\ref{tab:psf}.

{\King} presents the radially extended {\PSFK} that continues out to $r\simeq5^{\circ}$; this PSF is a composite of his own measurements with the 48" telescope at the Palomar Observatory and, amongst other, the measurements of \citet{Va:58}, who used the 21" reflector at the Lowell Observatory. \citet{Ko:73} presents {\PSFKo} for the radial range $7\arcsec\le r\le1^{\circ}$, which was also measured with the 48" telescope at Palomar. With {\PSFC}, {\CV} present the, so far, radially most extended PSF, which reaches $r=90^{\circ}$ and covers more than $30\,$mag in intensity. When {\CV} measure {\PSFC}, they simplify their analysis and assume that the PSF is well fitted by a sum of three Gaussian profiles \citep[this approach is earlier also used by][]{Va:48,Br:78}. Their analysis on the origin of the PSF and various sources of errors is thorough, but it is applied to NGC\,3379, which is an extended elliptical galaxy with a slowly decreasing surface brightness; they show that in this case the PSF is of minor importance. {\PSFCa} and {\PSFCb}, which only differ in the core, are claimed to be measured at a seeing of $0\farcs5$ and $1\farcs0$, respectively, instead of the real values $1\farcs0$ and $2\farcs0$ that can be measured in the profiles.

\citet{SaScRo.:08} measure a spectroscopy PSF using the lens array integral field unit of the Potsdam Multi Aperture Spectrograph (PMAS). The published PSF was measured using saturated data, which resulted in a too wide core profile. Here, {\PSFPa} is presented, where the core profile is replaced with the PSF of a faint star that was observed at $0\farcs8$ seeing (this procedure causes the kink in {\PSFPa} that is seen at $r\approx1\farcs5$). {\PSFPa} only extends to $r\simeq25\arcsec$, but it agrees well with the imaging PSFs -- this indicates that PSF effects are plausibly equally critical in imaging and spectroscopy data. \citet{MoRoScSt:06} show the only other example of an extended spectroscopy PSF that I know of, for the VIsual Multi-Object Spectrograph at the Very Large Telescope, but it only reaches $r=12\arcsec$. Moreover, {\Slater} study the instrumental origin of the PSF and present the composite {\PSFS}, which extends to $r\simeq64\arcmin$. These authors show that the aureole brightness depends strongly on where measurements are made in the field. Measurements that are offset from the optical axis may, depending on the amplitude of the offset and changing optical conditions, result in asymmetric PSFs.

The PSF studies that were mentioned so far do not examine any time or wavelength variations outside the centre PSF. Such variations are since long found in photometer measurements near the solar limb \citep{Mi:53,Pi:54} and off the bright limb of the Moon \citep{Pi:73}; the latter study presents three sections of the aureole with {\PSFPi}. \citeauthor{Pi:73}, moreover, finds a correlation between the measured aureole and times of mirror re-aluminizing and washing, as well as the telescope location. The aureole that he measures with data from the McDonald Observatory (not published) is time invariant, whilst the one that he measures at the Goethe Link Observatory increases by up to a factor ten within a few months. \citet{Mi:02} measures inner aureoles at the Haute-Provence Observatory with various imaging filters that also show temporal changes. He presents the only published radially extended PSFs that are available for two bands at two distinct times (they are separated by three months; \PSFVa, \PSFVb, \PSFIa, and \PSFIb). The two $V$-band PSFs differ by up to $1\,\magg$, whilst the two $i$-band PSFs show a smaller difference. In general, PSFs show weak variations with wavelength, except measurements in the $i$-band that in some parts are affected by the CCD-specific red halo effect \citep{SiClHa.:98}. \PSFK, \PSFKo, \PSFPi, and {\PSFC} (as well as PSF$_{\text{S74}}$, see Sect.~\ref{sec:Bernstein}) were measured with photographic and photomultiplier data, and more recent PSFs with CCDs.

Several studies stack a large number of images of the Sloan Digital Sky Survey (SDSS) and then measure one PSF per band; for example, \citealt{ZiWhBr:04}, \Jong, \citealt{BeZaCa:10}, and \citealt{TaDo:11}, which extend to $r\simeq30$--$60\arcsec$. The $g$- and $i$-band SDSS PSFs are shown in Fig.~\ref{fig:psf} versus $10^3\times r$, separated from the other PSFs. {\Jong} extrapolates his PSF for $r>71\arcsec$, using a $r^{-2.6}$ power law (dotted dark green line in Fig.~\ref{fig:psf}). This extrapolation is poorly justified in comparison with, for example, {\PSFK} and \PSFS, which decline as $r^{-2}$, and other PSFs are even more shallow [\PSFPi ($\circ$, $\bullet$), \PSFKo, and \PSFC]. The SDSS PSFs are all averaged using a large number of 53\,s exposures that were collected during years of observations. Such PSFs, of individual filters, should be nearly identical, if they are time invariable and differences across the field are ignored. Figure~\ref{fig:psf} shows that the $i$-band filter PSFs differ by up to two magnitudes, in particular the $i$-band PSF of \citet{TaDo:11} lies above the other ones. The large deviation between the PSFs in the outermost parts illustrates that the SDSS exposure time of 53\,s might be too short for these measurements. It is plausible that the SDSS PSFs are affected by either temporal variations, optical-path differences across the field, or both.

Two of the three remaining PSFs in Fig.~\ref{fig:psf} are {\PSFMBH} and {\PSFMBHT}. {\MBH} present {\PSFMBH} in their analysis of NGC 5907 and {\PSFMBHT} is derived here, both PSFs are discussed further in Sect.~\ref{sec:ngc5907}.

Most of the PSFs mentioned here are measured with telescopes that use reflective mirrors; remaining parts use scaled intensity estimates of the sky away from the sun. \citet{AbDo:14} instead use an optical configuration with eight, comparatively small, refracting telephoto lenses. Their resulting {\PSFAbr} lies markedly lower than all other PSFs where $r\ga30\arcsec$.

The presented PSFs clearly illustrate that there is no empirical support for truncating a PSF at a shorter radius. Except in the innermost seeing-dependent core, and where $100\la r\la400\arcsec$, {\PSFK} appears to present an approximative lower limit of the extent of scattered light throughout the radial range in reflective telescopes. Except {\PSFS} and {\PSFAbr}, the outer PSF (where $r\ga500\arcsec$) was measured once in the early eighties (\PSFC), and only thrice before that (\PSFK, \PSFKo, and \PSFPi). Additionally, \citet{WuBuDe.:02} claim that they measured a PSF that extends to $r=1700\arcsec$, but they do not present it (and the data are lost, Wu, priv.\ comm.). I speculate that the $r^{-2}$ dependence at large radii of the other PSFs, at least partially, occurs due to an obstructed pupil and reflective surfaces, whilst PSFs that are even brighter occur due to the degradation of, as well as deposition of dust, on reflective optical surfaces.

\subsection{Further evidence against a steeper than $r^{-2}$ power-law slope that contradicts the conclusion of {\Bernstein}}\label{sec:Bernstein}
{\Bernstein} discusses PSFs from the viewpoint of their slope at different radii. She notes that the slopes of the outer PSF differ, which are measured with the 48" telescope at Palomar by three different authors. To this purpose, she compares the $r^{-2}$ slope of {\King} for $r<228\arcsec$ (the exact dates of these measurements are not specified) with the $r^{-1.7}$ slope of \citet[measured August 1971--April 1972]{Ko:73} for $3\le r\le30\arcmin$, and the $r^{-2.6}$ slope of \citet[measured March--April 1971]{Sh:74} for $r=100\arcsec$, and summarizes that the PSF of this telescope is not well determined within a time frame of a few months. She notes that differences are probably the result of filter and emulsion use, mirror cleanliness, and measurement errors. In view of the two sets of PSFs of \citet{Mi:02}, it seems plausible that the temporal resolution of the Palomar PSFs is too poor to draw any conclusion about how the PSFs change with time. Note, however, that the feature at $r\approx280\arcsec$ is seen in both {\PSFKo} and PSF$_{\text{S74}}$.

The comparison of slopes that {\Bernstein} presents appears less dramatic when all PSFs are collected in the same plot. All but one PSF of {\Bernstein} are shown in Fig.~\ref{fig:bpsf}. The exception are the data of \citet{MiClWa:89}, who use {\PSFK} and present data of the small, but extended, planetary nebula BD+30\degr3639 (in their fig.~1); {\Bernstein} seemingly uses these data as a PSF. {\Bernstein} measures {\PSFB} that reaches $r\simeq400\arcsec$, and is centred on the optical axis. She also measures a second PSF that is offset by $3\arcmin$ from the optical axis; in view of the lack of details regarding the optical setup, and with respect to the discussion and results of \Slater, it is uncertain how the offset PSF can be compared to \PSFB. For example, on which side of the optical axis is the offset PSF measured? Whilst there are local deviations from a slope that decreases as $r^{-2}$, there is no evidence for a globally steeper slope in telescopes that use reflective mirrors, neither in Fig.~\ref{fig:bpsf} nor in Fig.~\ref{fig:psf}. {\PSFB} is partly lower in the outermost regions, where it is also very noisy.

\subsection{Extrapolating and normalizing the PSFs}\label{sec:enpsf}
I chose to normalize all PSFs that are measured near the centre to $r=90^{\circ}$, which is the largest radius that was considered in the derivation of a PSF (see the information on \PSFC). Even larger radii cannot be excluded. In the centre region, I extrapolated each PSF to $r=0$ with centred and fitted Gaussian profiles. All details of the normalization and extrapolation to larger radii of individual PSFs are given in Sect.~\ref{sec:aenpsf}.

The outer regions of the PSFs are only poorly known, at best, which makes the normalization uncertain. In this paper, I delimited the study of objects to the radial range $r<450\arcsec$, and use PSFs in the radial range $r<900\arcsec$ (cf.\ Sect.~\ref{sec:toy}).

\begin{figure}
\centering
\includegraphics[width=\columnwidth]{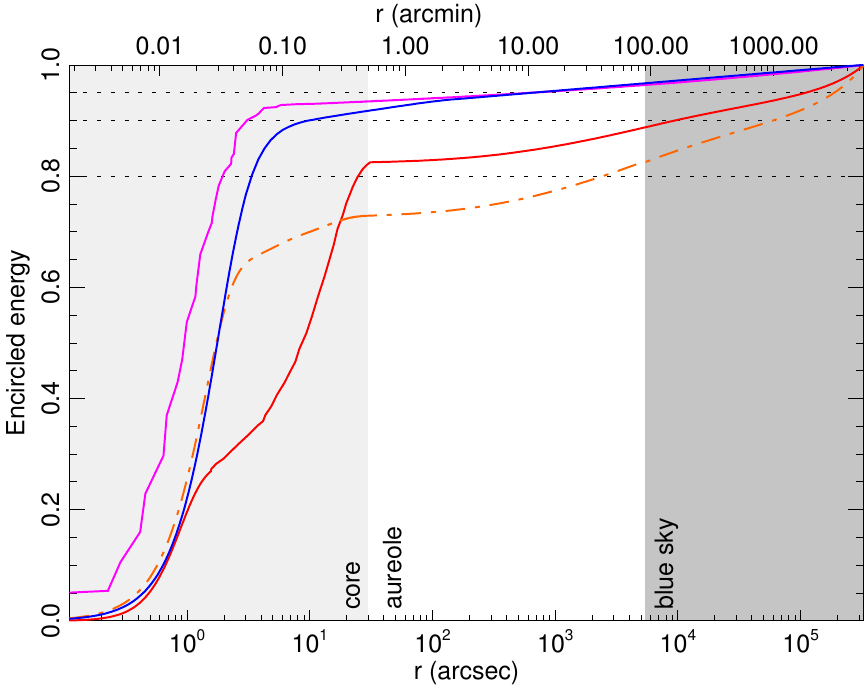}
\caption{Encircled energy versus the radius $r$. The four shown PSFs are: {\PSFK} (magenta, uppermost line), {\PSFCa} (red, lowermost line for $r<20\arcsec$), {\PSFCb} (orange, lowermost dash-dotted line for $r>20\arcsec$), and {\PSFVa} (blue). The light (medium) grey region indicates the PSF core (blue sky), and the white region the aureole, as defined for {\PSFCa} and {\PSFCb} (\CV). The horizontal dotted lines at $0.8$, $0.9$, and $0.95$ are guides.}
\label{fig:ipsf}
\end{figure}

\subsection{Measuring the encircled energy}\label{sec:eepsf}
Whilst the radially extended PSF may reach $r=90^{\circ}$, and beyond, only the inner parts, which are induced by the instrument and not the atmosphere, are expected to affect the accuracy of flux measurements (\CV; \Bernstein). {\CV} measure the $B$-band {\PSFC} and find that the energy fraction in the aureole and the blue sky, 30 per cent, corresponds to the atmospheric extinction in the $B$-band. To check how general this finding is, the encircled energy is shown versus radius for four PSFs in Fig.~\ref{fig:ipsf}: {\PSFCa} and {\PSFCb}, which share the outer profile, and {\PSFK} and {\PSFVa}. About 28 per cent of the encircled energy is contained in the aureole and the blue sky parts of {\PSFCb}, in agreement with {\CV}. However, the value is about 18 (6.5; 8) per cent for {\PSFCa} ($B$-band {\PSFK}; $V$-band {\PSFVa}). Also, the outer parts of {\PSFC} were measured without any filter (Table~\ref{tab:psf}). {\PSFC} (and thereby {\PSFCa} and \PSFCb) becomes more shallow for $r\ga1000\arcmin$, possibly as a result of fitting the measurements with three Gaussian profiles. The shallower PSF causes slopes that are steeper than those of {\PSFK} and {\PSFVa} in the outermost (blue-sky) region, cf.\ Fig.~\ref{fig:ipsf}.

Considering the quoted percentages and that the extinction decreases for redder wavelengths, with the currently measured PSFs there is clearly no simple correlation between the fraction of energy in the outer PSF and the atmospheric extinction.

\subsection{Recommendations for new measurements of the PSF}\label{sec:newpsf}
The total number of measured PSFs that extend beyond a few minutes of arc is small; \PSFK, \PSFKo, \PSFPi, \PSFC, and {\PSFS} extend this far. Only {\PSFS} was measured recently. Neither a wavelength dependence nor temporal variations were considered in the measurements of the outer parts of these PSFs; the general application of these static PSFs to deconvolve data is discouraged. It is also important that all flux is included in the calculation of azimuthally averaged profiles; the PSF is somewhat underestimated, if, for example, diffraction spikes are masked (this is measured here for the case of NGC 5907, cf.\ Sect.~\ref{sec:ngc5907newanalysis}).

Nowadays, PSFs are exclusively measured using combined data of faint and bright stars. Measurements in different radial regions are combined by matching them in overlapping regions. Faint stars are used to measure the seeing-dependent core, and bright stars are used to determine the fainter outer regions. Very bright stars are, unfortunately, limited in number. Even if a bright star is used, the surface brightness cannot be accurately measured beyond some radius with reasonable exposure times (say, shorter than about $1800\,$s).

\begin{table*}
\caption{Properties of the Sun, the Moon, and the brightest planets and stars on the sky.}
\centering
\label{tab:ppsf}
\begin{tabular}{lrrrrccl}\hline\hline\\[-1.8ex]
Object & \multicolumn{2}{c}{Diameter} & \multicolumn{2}{c}{$m_V$ (mag)} & \multicolumn{2}{c}{$r_{\text{PSF}}$ / Decl.} & Comment\\
       & apogee & perigee & \multicolumn{1}{c}{max} & \multicolumn{1}{c}{min} & apogee & perigee &\\\hline\\[-1.8ex]
\object{the Sun}  & $31\farcm6$ & $32\farcm7$ & \multicolumn{2}{c}{$-26.45$} & $50\parcmin$ & $\phantom{0}52\parcmin$ \\
\object{the Moon} & $29\arcmin$ & $34\arcmin$ & $-2.5\phantom{0}$ & $-12.9\phantom{0}$ & $47\parcmin$ & $\phantom{0}54\parcmin$ & The full Moon\\
\object{Venus} & $9\farcs7$ & $66\arcsec$   & $-3.8\phantom{0}$ & $-4.9\phantom{0}$ & $17\arcsec$ & $130\arcsec$ & Crescent at perigee\\
\object{Mars}  & $3\farcs5$ & $25\farcs1$ & $+1.6\phantom{0}$  & $-3.0\phantom{0}$ & $13\arcsec$ & $\phantom{0}40\arcsec$ &\\
\object{Jupiter} & $29\farcs8$ & $50\farcs1$ & $-1.6\phantom{0}$ & $-2.94$ & $43\arcsec$ & $\phantom{0}88\arcsec$ & \\
\object{Saturn}  & $14\farcs5$ & $20\farcs1$ & $+1.47$ & $-0.24$ & $13\arcsec$ & $\phantom{0}36\arcsec$ & Ignoring the rings.\\
\object{$\alpha$CMa} &               &               & \multicolumn{2}{c}{$-1.47$} & \multicolumn{2}{c}{$-16^\circ42\arcmin58\arcsec$} & Sirius A\\
\object{$\alpha$Car} &               &               & \multicolumn{2}{c}{$-0.72$} & \multicolumn{2}{c}{$-52^\circ41\arcmin44\arcsec$} & Canopus\\
\object{$\alpha$Cen} &               &               & \multicolumn{2}{c}{$-0.27$} & \multicolumn{2}{c}{$-60^\circ50\arcmin02\arcsec$} & Alpha Centauri A+B\\
\object{$\alpha$Boo} &               &               & \multicolumn{2}{c}{$-0.04$} & \multicolumn{2}{c}{$+19^\circ10\arcmin56\arcsec$} & Arcturus\\
\object{$\alpha$Lyr} &               &               & \multicolumn{2}{c}{$+0.03$} & \multicolumn{2}{c}{$+38^\circ47\arcmin01\arcsec$} & Vega\\
\object{$\alpha$Aur} &               &               & \multicolumn{2}{c}{$+0.03$--$0.16$} & \multicolumn{2}{c}{$+45^\circ59\arcmin53\arcsec$} & Capella\\
\hline
\end{tabular}
\tablefoot{Column.~1, object name; Cols.~2 and 3, object diameter at apogee and perigee; Cols.~4 and 5, minimum and maximum $V$-band magnitude for planets and solar system bodies, and total magnitude for the stars; Cols.~6 and 7, minimum and maximum values on $r_{\text{PSF}}$ using {\PSFVa} for the Sun, the Moon, and the planets, and declination for the stars; and Col.~8, additional comments. Only approximate values on the diameters, magnitudes, and coordinates are presented here (see any book on planetary science).}
\end{table*}

\begin{figure}
\centering
\includegraphics[width=\columnwidth]{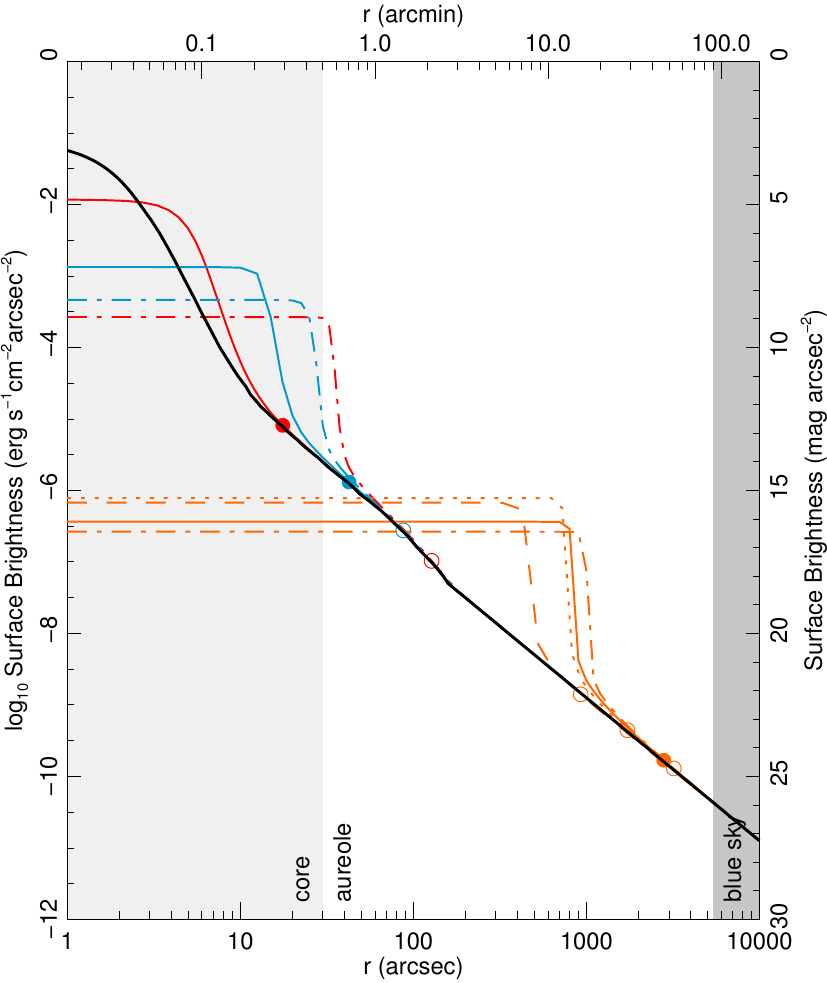}
\caption{Comparison of {\PSFVa} (black line) with surface-brightness profile cuts of Venus (red lines), Jupiter (blue lines), the full Moon (orange solid and dash-dotted lines), and the half Moon on the symmetry axis (orange dashed line) and the axis orthogonal to that (orange dotted line). Profiles of objects at apogee (perigee) are drawn with solid (dash-dotted) lines. All profiles, but {\PSFVa}, were divided with the illuminated area of the respective object. The light (medium) grey region indicates the PSF core (blue sky), and the white region the aureole. Bullets and circles indicate $r_{\text{PSF}}$ of individual profiles, cf.~Table~\ref{tab:ppsf}.}
\label{fig:ppsf}
\end{figure}

Earlier PSF studies also make use of observations of Jupiter, the Moon, and even the Sun. Compared to centres of stars, surface brightnesses of the planets and the Moon are lower, but integrated intensities can be drastically higher.

Extended objects are not directly comparable to point sources. However, they appear as point sources, beginning at a geometry-dependent angular distance $r_{\text{PSF}}$, after they are normalized with the illuminated surface area (including a correction for intensity variations across the surface). To illustrate this, I made the simplifying assumption that each extended object is a disc of specified diameter with a constant (homogeneous) brightness. Each surface-brightness structure was convolved with {\PSFVa} and the result was divided with the illuminated area. With the smaller planets (the larger planets; the Moon and the Sun) I created model images where I used pixels that are $0\farcs55$ ($2\farcs5$; $10\arcsec$) on the side. Surface-brightness profile cuts at the apogee and the perigee of Venus, Jupiter, and the Moon are compared with {\PSFVa} in Fig.~\ref{fig:ppsf}. 

Object diameters at perigee and apogee, as well as magnitudes of the Sun, the Moon, the brighter planets, and the six brightest stars are collected in Table~\ref{tab:ppsf}. Approximate values on magnitudes and sizes should suffice in this context -- real PSFs are preferably derived by matching overlapping PSFs. The beginning PSF radius $r_{\text{PSF}}$, where the surface brightness of each respective extended object deviates by less than 5 per cent from {\PSFVa}, is shown in the same table. Adopting a generous margin, all objects appear as PSFs outside an angular radius of two times the respective object diameter, except Mars at apogee ($3.7\times$ the diameter). $r_{\text{PSF}}$ is shown in Fig.~\ref{fig:ppsf} also for the half Moon at apogee, as viewed on the symmetry axis and the axis that is orthogonal to that; the corresponding beginning PSF radii are $r_{\text{PSF}}=16\arcmin$ and $r_{\text{PSF}}=29\arcmin$. It seems appropriate to observe other bright objects instead of the faint Mars and Saturn at their apogee, but in particular Mars is significantly brighter at perigee.

The intermediate PSF range, say $1\arcmin\la r\la1^{\circ}$, is well measured with Mars and Saturn at perigee, and with Venus and Jupiter at any time. De Vaucouleurs (\citeyear{Va:58}) uses Jupiter to measure a PSF out to $r=5^{\circ}$, $29\,$mag below the integrated magnitude (these measurements are part of \PSFK, cf.\ Table~\ref{tab:psf}). {\Bernstein} raises some concern that the faintest intensities measured around Jupiter are affected by Zodiacal light, but Jupiter is bright and the slope agrees with the other measurements. Venus is a crescent at perigee, the beginning PSF radius is therefore slightly smaller than what I calculated using a disc.

The outer PSF, say $1\la r\la10\degr$, is well measured using the bright Moon, say, when half or more of the Moon is illuminated. \citet[and references therein]{KrSc:91} provide a detailed account for issues related to measurements of the surface-brightness due to the Moon across the sky \citep[also see][]{Pa:03}. The outermost parts of the PSF, $1\la r\la90\degr$ (and beyond), could be estimated using the Sun (\citealt{Hu:48}, \CV). The illuminated and projected area of the Moon varies between the new Moon and the full Moon -- the angular distance should be measured from the centre of mass of the illuminated area.

The edge-on galaxy models that are presented in Sect.~\ref{sec:toy} indicate that it is necessary to measure the PSF out to a radius that is 1.5 times larger than the measurements, to ensure that observed data can be corrected for integrated scattered light. Current measurements of PSFs are crude and few, which is why it is difficult to judge their influence on observations in general. New measurements of radially extended PSF -- as function of angle, time, wavelength, and position in the field -- are needed for all telescope and instrument setups that are used to observe extended objects. There are several attempts to model the outer PSF theoretically, but it is not understood how the inverse-square decline with radius is created.

Lots of observing time would be required, if it was demanded that each project observes individual PSFs. With the \textit{HST} PSF model code \textsc{Tiny Tim} (see the discussion on space-based PSFs in Sect.~\ref{sec:HSTpsf}) as an example, it seems worthwhile to explore possibilities to develop a similar tool for ground-based telescopes. Analysis work of both past and new observations could be improved if there was a PSF lookup library for different instrument modes at a telescope, as function of filter or wavelength, time, and position within the observable field. New observations could at first be made at some weekly or monthly intervals to determine the temporal variations of the PSF; it might be a good idea to also record the time since the last mirror aluminization and washing, the air quality, ground reflective properties, and the like. At the moment it seems that the extended PSF must be individually measured for each observed exposure, but the suggested approach would provide good foundations and constraints to develop the theory of the outer parts of the PSF.

\section{Method and its application to example models}\label{sec:methods}
The simulations of the surface-brightness structure of the example galaxies and NGC 5907 were split into three steps. I selected a set of measured PSFs that can be used to estimate varying scattered-light effects. Thereafter, I configured models of the surface-brightness structure. Finally, I applied the PSFs to the model structures and analysed the outcome. These steps are described in the following three subsections. I apply the method to three example models in Sect.~\ref{sec:toy}.

\subsection{Choosing representative PSFs}
Representative PSFs should describe both temporal variations and the red-halo effect in the $i$ band, and they should extend out to $r\simeq900\arcsec$, which is twice the maximum object radius that I consider. I chose to use the only published extended PSFs that were measured with both the (Cousins) $V$ and the (Gunn) $i$ bands, at two distinct occasions that were separated by three months: {\PSFVa} and {\PSFIa} that were measured three months before {\PSFVb} and {\PSFIb}; the difference with radius of {\PSFVb}$-${\PSFVa} is on average higher than that of {\PSFIb}$-${\PSFIa} (Fig.~\ref{fig:psf}). The comparison of the PSFs in Figs.~\ref{fig:psf} and \ref{fig:bpsf} demonstrate that {\PSFVa} is perhaps $\la0.5\,\magg$ brighter at intermediate radii than the other PSFs; the value is a bit uncertain since it was necessary to scale several PSFs. At larger radii, $r\ga80\arcsec$, it is difficult to draw such a conclusion. I used {\PSFVa} as an indicator of average PSF effects with the example models in Sect.~\ref{sec:toy}. {\PSFVb} is much brighter, and is more representative of an upper limit. (See Sect.~\ref{sec:ngc5907newanalysis} for a comparison of {\PSFMBH} and \PSFMBHT.) The radially extended $B$-band {\PSFK} is used as a lower limit indicator of scattered light, both with the example models and with NGC 5907. Moreover, {\PSFIa} represents the $i$-band average SDSS PSFs well where $r\ga10\arcsec$. {\PSFIb} is used as an upper limit in the $i$ band, similar to how {\PSFVb} is used. Some PSFs show more light at large radii than the $r^{-2}$ power-law of {\PSFK} (see Sect.~\ref{sec:PSF}); the predictive ability at larger radii is therefore weaker.

In the simulations, I assumed that the two $V$-band {\PSFVa} and {\PSFVb}, as well as the $B$-band {\PSFK}, are the same in the $R$ band -- I do this under the assumption that measured PSFs of different bands (except the $i$ band) are very similar, cf.\ Sect.~\ref{sec:lpsf}. The colour predictability is delimited to $R-i$. I did not differentiate between Cousins, Harris, or other photometric systems. All five PSFs were measured at a seeing of several arc seconds, which is why the spatial resolution in the brightest centre region is poor. Here, faint diffuse emission is studied, where the spatial resolution and resulting lower intensities in the centre regions are of minor importance.

\subsection{Setting up sets of model surface-brightness structures}
Two-dimensional surface-brightness structures of edge-on disc galaxies are suitably described in cylindrical coordinates. The space-luminosity density $I'$ can be described by \citep{Kr:88,KrSe:81a}
\begin{eqnarray}
I^{\prime}(r',z')=I^{\prime}_{0,0}\exp\left(-\frac{r'}{h_{\text{r}}}\right)\times2^{-2/n_{\text{s}}}\text{sech}^{2/n_{\text{s}}}\left(\frac{n_{\text{s}}z'}{z_{0}}\right),
\end{eqnarray}
where $I^{\prime}_{0,0}$ is the centre intensity, $r'$ the radius, $h_{\text{r}}$ the scale length, $z'$ the vertical distance from the centre, $z_{0}$ the vertical scale height, and $n_{\text{s}}$ is set to $1$, $2$, or $\infty$. The intensity drops to zero at a galaxy-specific truncation radius. When a disc galaxy is projected edge on, and the truncation radius as well as dust extinction are ignored, the surface-brightness structure $I(r,z)$ becomes
\begin{eqnarray}
I(r,z)=I_{0,0}\frac{r}{h_{\text{r}}}K_{1}\left(\frac{r}{h_{\text{r}}}\right)\times2^{-2/n_{\text{s}}}\text{sech}^{2/n_{\text{s}}}\left(\frac{n_{\text{s}}z}{z_{0}}\right),
\end{eqnarray}
where $I_{0,0}=2h_{\text{r}}I^{\prime}_{0,0}$ is the centre intensity, $r$ the major-axis radius, $z$ the minor-axis (vertical) distance from the centre of the disc, and $K_{1}$ the modified Bessel function of the second kind. The intensity is slightly lower at larger radii when the truncation radius is finite and taken into account; considering how uncertain the PSFs are, this effect is ignored here. I assume in Sect.~\ref{sec:toy} that the disc is isothermal ($n_{\text{s}}=2$), and in Sect.~\ref{sec:ngc5907} it seems to work better using $n_{\text{s}}=1$ with NGC 5907. No attempt is otherwise made to make a perfect fit of the centre region of NGC 5907. The relation $\mu=-2.5\log_{10}(I)$ is used to convert between magnitudes $\mu$ and intensities $I$.

\subsection{Models and measurements analysis procedure}
Each two-dimensional $R$-band model image is convolved individually with the three resampled and normalized two-dimensional $V$-band {\PSFVa} and {\PSFVb}, and the $B$-band {\PSFK}. This is repeated for each $i$-band model image with {\PSFIa} and {\PSFIb}. The $i$-band models use the same model parameters as the $R$-band models, with one exception, $z_{0,i}=0.95z_{0,R}$, which results in negative slopes in $R-i$ with increasing vertical distance. Thereby, any red-excess haloes in convolved models are induced by the PSF. $\mu_{0,0,R}-\mu_{0,0,i}$ is set to $1.3\,\magg$, which produces a rough agreement with the $V-i$ profile for NGC 5907 of \citet[hereafter \Lequeux]{LeFoDa.:96}\footnote{The exact difference is unimportant to the demonstration here, where the main point is to illustrate the spatial dependence of the colour profile. All colour profiles are offset by the same value.}. The used PSF image is twice as large as the model image, to avoid PSF truncation effects in convolved images (see below). The PSF and the model images are resampled to use the same pitch and about 100--200 pixels on the side, typically, to keep calculation times short. All convolutions are made by direct integration.

Convolved surface-brightness profiles are plotted together with model surface-brightness profiles for a cut along the minor axis, directed outwards from the centre on the same (vertical) axis. $V$- and $R$-band measurements of NGC 5907 are plotted as well in Sect.~\ref{sec:ngc5907}; I used \textsc{dexter}\footnote{\textsc{dexter} is available at: \href{http://dexter.sourceforge.net}{http://dexter.sourceforge.net}.} \citep{DeAcEi.:01} to extract the data. The three resulting $R-i$ colour profiles are shown in a separate lower panel: the model, the convolved models that used the earlier {\PSFVa} and {\PSFIa}, and the convolved models that used the later {\PSFVb} and {\PSFIb}.

The PSF induces scattered light, where surface-brightness profile cuts of input models and convolved models differ. Larger differences between profile cuts of convolved models that use {\PSFVa}, {\PSFVb}, and {\PSFK} in the $R$ band, or {\PSFIa} and {\PSFIb} in the $i$ band, are indicators of stronger time dependence in the PSFs. The opposite applies when such differences are smaller. Measured values that are due to scattered light should fall on top of profile cuts of a model that is convolved with the PSF at the time of the observations.

A scattered-light halo radius $r_{110}$ is defined as the smaller limiting radius where the convolved model intensity at all larger radii is $\ge10$ per cent higher than the model intensity. $r_{110}$ depends on both object parameters and the PSF.

\begin{table}
\caption{Example model parameters.}
\centering
\label{tab:toypsf}
\begin{tabular}{lllll}\hline\hline\\[-1.8ex]
  & \multicolumn{1}{c}{$h_{\text{r}}$} & \multicolumn{1}{c}{$z_{0}$} & \multicolumn{1}{c}{$r_{\text{tr},1}$} & \multicolumn{1}{c}{$r_{\text{tr},2}$}\\\hline\\[-1.8ex]
small galaxy   & $\phantom{00}7\farcs5$ & $\phantom{0}1\farcs5$ & $\phantom{0}4\farcs0$ & $\phantom{0}10\arcsec$\\
medium galaxy & $\phantom{0}20\arcsec$ & $\phantom{0}4\farcs0$ & $10\arcsec$ & $\phantom{0}25\arcsec$\\
large galaxy       & $240\arcsec$ & $40\arcsec$  & $10\arcsec$ & $250\arcsec$\\[0.3ex]
\hline
\end{tabular}
\tablefoot{Column~2, major-axis scale length; Col.~3, minor-axis scale height; Col.~4, first truncation radius; and Col.~5, second truncation radius. The parameter $n_{\text{s}}=1$ in each case.}
\end{table}

\begin{figure*}
\centering
\includegraphics{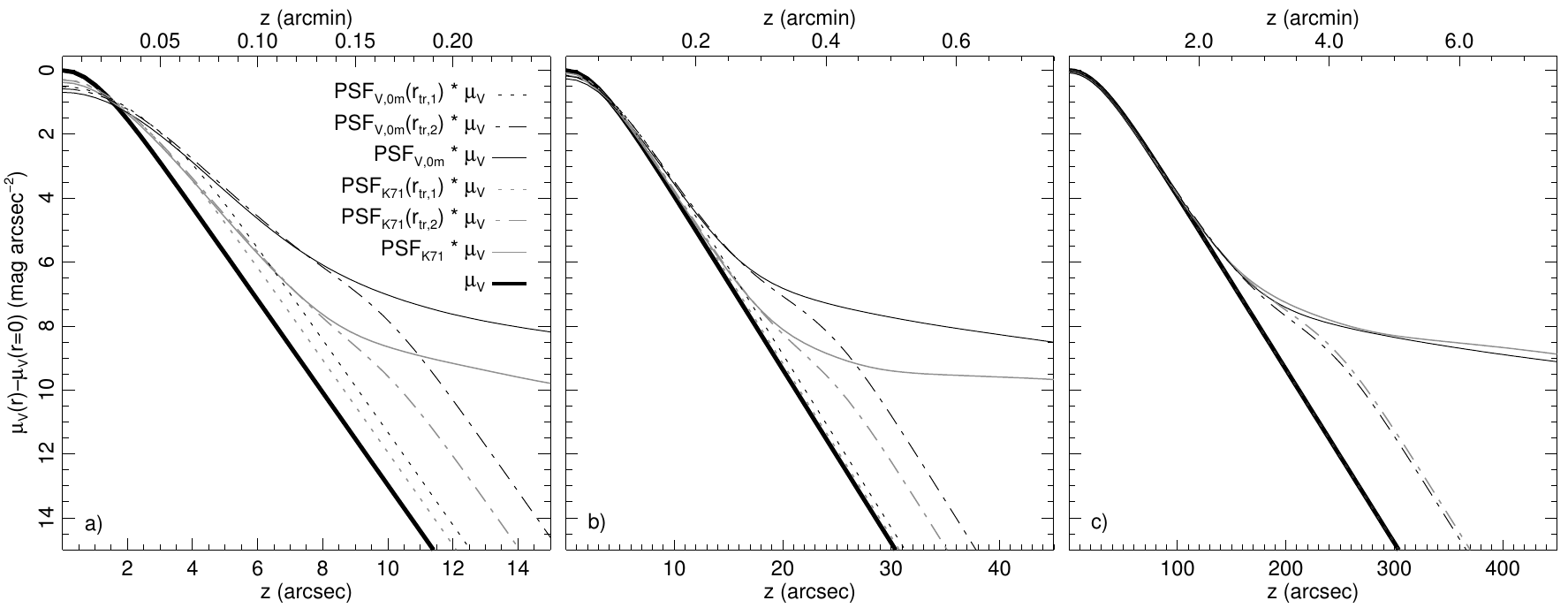}
\caption{Vertical-axis surface-brightness profiles that illustrate effects of radial truncation of {\PSFK} and {\PSFVa} in example models. The three panels show: \textbf{a}) a small edge-on galaxy, \textbf{b}) an edge-on galaxy of intermediate size, and \textbf{c}) a large edge-on galaxy. In each panel, the disc-galaxy model profile is drawn with a thick black solid line, and model profiles that were convolved with {\PSFVa} (\PSFK) are drawn with thin black (grey) lines. The profile of the model that is convolved with the complete PSF is drawn with a thin solid line. Profiles are also shown where each PSF was truncated at $r_{\text{tr,1}}$ ($r_{\text{tr,2}}$) with a dotted line (dash-dotted line) in each panel; \textbf{a}) $r_{\text{tr,1}}=4\arcsec$ and $r_{\text{tr,2}}=10\arcsec$; \textbf{b}) $r_{\text{tr,1}}=10\arcsec$ and $r_{\text{tr,2}}=25\arcsec$; \textbf{c}) $r_{\text{tr,1}}=10\arcsec$ (lines fall on top of the model line and are not visible) and $r_{\text{tr,2}}=250\arcsec$ (Table~\ref{tab:toypsf}).}
\label{fig:toypsf}
\end{figure*}

\subsection{Application of the method on three example models}\label{sec:toy}
I calculated three example models of a small, an intermediate, and a large galaxy that are viewed edge on, at a high inclination angle, to illustrate consequences of using radially truncated PSFs. I varied the minor-axis scale height $z_{0}$ and set $h_{\text{r}}=5z_{0}$ and $n_{\text{s}}=2$. Each model was convolved with {\PSFK} and an extrapolated {\PSFVa}. Additionally, each model was convolved with either PSF, after the PSF was truncated at a smaller radius $r_{\text{tr},1}$ or a larger radius $r_{\text{tr},2}$. Model parameters and truncation radii $r_{\text{tr}}$ are given for each model in Table~\ref{tab:toypsf}. The resulting minor-axis surface-brightness profiles are shown in Fig.~\ref{fig:toypsf}. Values that are quoted in parentheses below used \PSFK, and the other values \PSFVa.

\subsubsection{The small galaxy}
The convolved profiles show excess scattered light at all distances $z\ga1\farcs7$ (Fig.~\ref{fig:toypsf}a), whilst less light is seen at shorter distances (this decrease is largely caused by the seeing-dominated part of the PSFs). The scattered-light halo of the convolved model is too faint for $z\ga4\arcsec$ ($z\ga4\farcs3$) when the PSF is truncated at $r_{\text{tr}}=4\arcsec$. The corresponding value for $r_{\text{tr}}=10\arcsec$ is $z\ga8\arcsec$ ($z\ga8\arcsec$). The surface-brightnesses that were calculated using the three versions of the truncated {\PSFVa} differ by up to $0.2\,\magg$ at the centre, due to the variations of the individual normalization of the PSFs. The correct amount of scattered light at $z=15\arcsec$ is only achieved when the PSF is not truncated within, say, $r\approx19\arcsec$ ($15\arcsec\times10\arcsec/8\arcsec$). Both {\PSFK} and {\PSFVa} are about $10\,\magg$ fainter at $r=15\arcsec$, compared to the centre. The model surface brightness is about $8.2$ ($6.6$) {\magg} fainter than the convolved structure at $z=12\arcsec$; the signal-to-noise (S/N) value that would be required to extract the intensity structure of the model structure is about 1900 (440). At $z=7\farcs5$, the corresponding values are $3.7\,\magg$ and S/N$\approx$30 ($2.1\,\magg$ and S/N$\approx$7).

\subsubsection{The intermediate-size galaxy}
There is less excess light compared to the small galaxy near the centre, Fig.~\ref{fig:toypsf}b. The convolved profiles show excess scattered light at all distances $z\ga5\arcsec$, and they overlap each other using either PSF for $z\la11\arcsec$. When the PSF is truncated at $r_{\text{tr}}=10\arcsec$, there is an excess due to scattered light of only about $0.5$ ($0.2$) {\magg} for $z\ga10\arcsec$. When the PSF is instead truncated at $r_{\text{tr}}=25\arcsec$, the convolved model becomes significantly fainter than the model that is convolved using the full PSF for $z\ga16\arcsec$ ($z\ga18\arcsec$). The correct amount of scattered light at, say, $z=40\arcsec$ is only achieved when PSF is not truncated within, say, $r=65\arcsec$ ($40\arcsec\times25\arcsec/16\arcsec$). Compared to the centre, the PSF is about $12$ ($14$) {\magg} fainter at $r=40\arcsec$. The model surface brightness is about $7.3$ ($5.6$) {\magg} fainter than the convolved structure at $z=30\arcsec$, and the required S/N$\approx$830 (S/N$\approx$170); at $z=20\arcsec$ the corresponding values are about $2.6\,\magg$ and S/N$\approx$11 ($1.4\,\magg$ and S/N$\approx$3.6).

\subsubsection{The large galaxy}
Using either of the radially complete PSFs, the convolved model shows significant excess that is due to scattered light, beginning at $z\approx120\arcsec$, Fig.~\ref{fig:toypsf}c. The halo is not reproduced at all when the PSF is truncated at $r_{\text{tr}}=10\arcsec$. The scattered-light halo of the convolved model is too faint for $z\ga160\arcsec$ when the PSF is truncated at $r=250\arcsec$. The PSF is about $20\,\magg$ fainter at $r=330\arcsec$ than at the centre (see Fig.~\ref{fig:psf}). To measure the modelled intensity at $z\simeq250\arcsec$, which is about $4.1\,\magg$ fainter than the scattered-light halo, the required S/N $\approx44$. The convolved profiles of the two PSFs nearly overlap since they are the same for $r>200\arcsec$. It is important to note that the surface-brightness of the model that is convolved with {\PSFK} in Fig.~\ref{fig:toypsf}c is brighter at large radii than in it is in Figs.~\ref{fig:toypsf}a and \ref{fig:toypsf}b; this is expected, because, compared to where $r\la30\arcsec$, the PSF is more shallow at larger radii.

\subsubsection{Summarizing the results of the example models}
The faintness of the PSF cannot be a limiting factor in accurate analyses of extended edge-on disc galaxies. It is instead the radial extent of the PSF that sets the limit on what can be corrected for. Not only small edge-on galaxies are strongly affected by scattered light, but also large galaxies, and time-dependent variations of the PSF cause variable structures.

The only way to remove scattered light, in the form of galaxy haloes, is through deconvolution with an accurately determined PSF and observations that, at least, cover all brighter regions at high enough signal-to-noise. However, the required accuracy in the measurements quickly becomes enormous with increasing vertical distances, and the required accuracy of the PSF cannot be lower than this, but is likely higher. It should, nevertheless, be safe to use a PSF with a radius that is twice as large as the maximum measured radius, accounting for symmetric objects. The tests above indicate that for these extended edge-on galaxies the minimum radial extent of the PSF is 1.5 times the outermost measured radius of the galaxy.

This study applies to already correctly determined surface-brightness structures and PSFs. Two factors that may play an important role to their determination are the assumed sky background level and scattered light from surrounding objects. Additionally, scattered light also affects the sky background. I show one example of how a slightly different value on the sky background changes the outcome of the analysis for observations of NGC 5907 in Sect.~\ref{sec:ngc5907}.

\section{On observations of the edge-on galaxy NGC 5907}\label{sec:ngc5907}
An extended faint halo was first found around an edge-on disc galaxy in $R$-band observations of NGC 5907 (\citealt{SaMoHaBo:94}; \MBH). Here, I re-examine the observations in the visual wavelength range of this galaxy with a focus on the role of integrated and time-varying scattered light. My analysis approach is to use both the originally derived {\PSFMBH} and vertical surface-brightness structure of \MBH, as well as {\PSFMBHT} and a surface-brightness structure that I derived myself as a consistency check, using the original data.

\subsection{Original observations and the derivation of \PSFMBH}\label{sec:ngc5907obs}
The first observations of the halo of NGC 5907 by {\MBH} were made using a Harris $R$-band filter with the Kitt Peak National Observatory No.~1 0.9m telescope on 1990 April 29 and 30, as part of a five-night observing run. The two nights were photometric and the seeing was $3\farcs5$. {\Lequeux} made follow-up observations of NGC 5907 using the $V$-band filter with the Canada-France-Hawaii Telescope at a seeing of $1\arcsec$. The telescope archive reveals that the observations were made on 1995 June 3 to 7. \citet[hereafter \Zheng]{ZhShSu.:99} observed NGC 5907 using a BATC 6660{\AA} filter with the Beijing Astronomical Observatory 0.6/0.9m telescope in $23$ nights on 1995 January 31 to June 27. The seeing was about $4\arcsec$. The filter bandwidth ratio of the $R$ band ($1380$\AA) to the BATC 6660{\AA} band ($480$\AA) is about $2.9$, which corresponds to a magnitude difference of $-1.15\,\magg$, assuming a constant intensity across the bandpass. The authors instead apply an offset of $0.3\,\magg$ to the $R$-band data of \MBH, to match them with their 6660{\AA}-band data -- the chosen value is not further motivated.

{\MBH} determine that the error in the 15 sky frames that they use is $1760$ ADU (analog-to-digital unit), and the associated error is $5.1$--$10.3$ ADU. They also derive {\PSFMBH}, using two brighter field stars, which extends out to $r=116\arcsec$; this PSF is shown in Fig.~\ref{fig:psf}. Outside the seeing disc, {\PSFMBH} lies above {\PSFVa} and {\PSFVb} for $r\la10\arcsec$, is closer to {\PSFVa} where $20\arcsec\la r\la70\arcsec$, and then turns upwards for larger radii. Additional details on the original observations of NGC 5907 are given in Sect.~\ref{sec:aongc5907}.

\subsection{Reconsidering the analysis of the original data}\label{sec:ngc5907newanalysis}
Two issues with the original study of {\MBH} motivate a reassessed analysis of the measurements: the background in the field around the galaxy does not appear flat on large scales, and the integrated PSF is underestimated when diffraction spikes are masked. I made a detailed reconsideration of the analysis of the originally reduced data of {\MBH}, cf.\ Sect.~\ref{sec:arngc5907}.

A flat background -- at the level of a few counts -- is important when measuring surface-brightnesses that are only slightly fainter than the sky. The new analysis revealed {\PSFMBHT} that is brighter at larger radii than {\PSFMBH}, see Fig.~\ref{fig:psf}; it is brighter because of a lower background value in the vicinity of the two saturated stars that are used to derive the PSF. The value is only $3$--$4\,$ADU lower than the value that {\MBH} seem to use, and is within the error bars of their flat-field image ($5.1$--$10.3\,$ADU). The new {\PSFMBHT} is also mostly within the error bars of {\PSFMBH}. The difference between {\PSFMBH} and {\PSFMBHT} is about $0.5\,\magg$ for $40\la r\la60\arcsec$.

The galaxy image is the result of the convolution with a PSF where diffraction spikes are present, therefore their contribution must not be removed from the PSF when it is used to correct object images for PSF effects. However, the difference due to masked diffraction spikes is small. A masked PSF is, when compared to an unmasked PSF, $0.1\,\magg$ fainter at $r\simeq20\arcsec$, this difference decreases to $0.0\,\magg$ at $r\simeq40\arcsec$.

\begin{figure*}
\sidecaption
\includegraphics{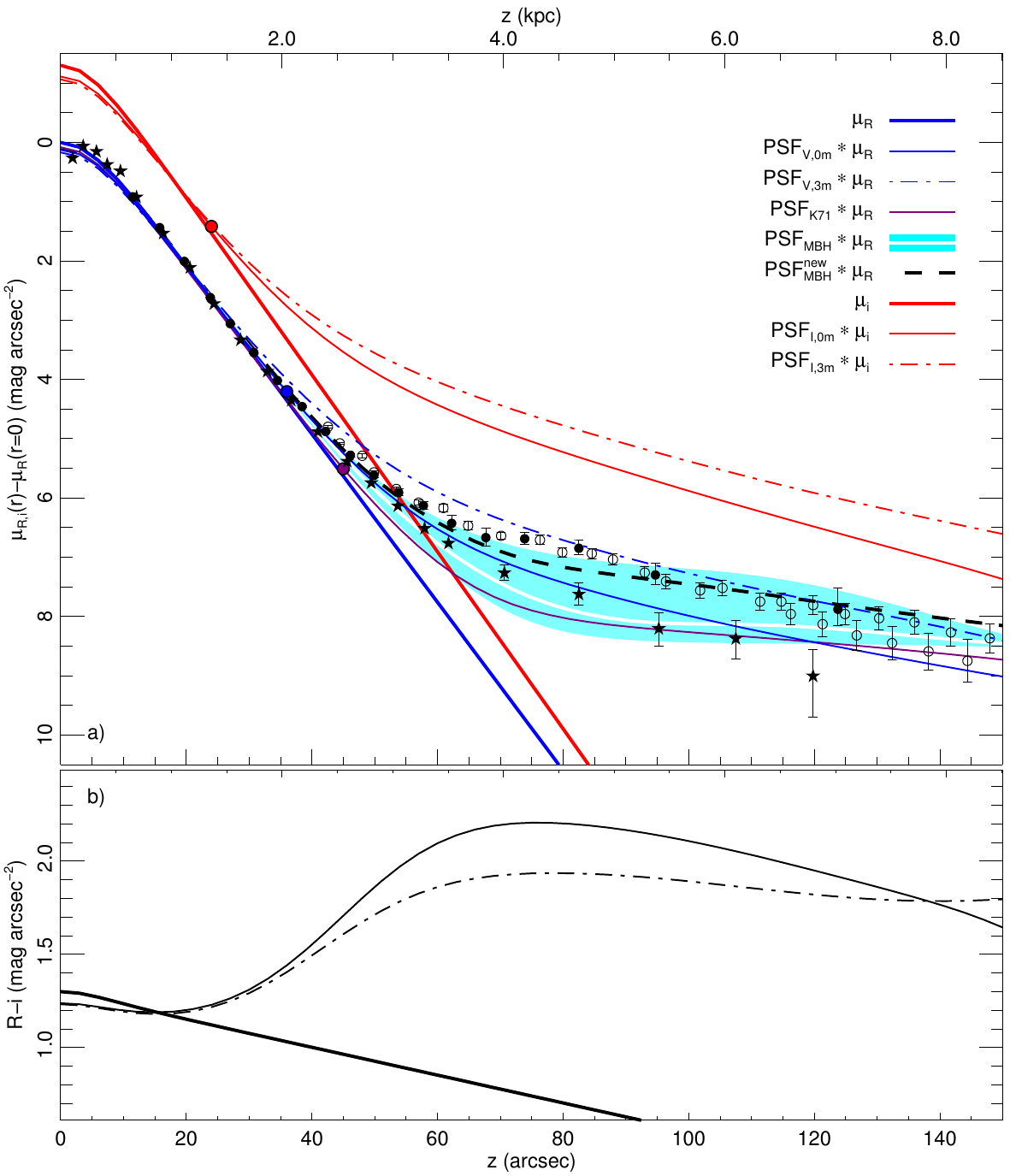}
\caption{Vertical-axis $R$-band and $i$-band surface-brightness profiles versus the vertical distance $z$ of models and measurements of the edge-on galaxy NGC 5907. \textbf{a}) Blue and purple lines show $R$-band profiles, and red lines $i$-band profiles. Model profiles are drawn with thick solid lines. Solid (dash dotted) lines are profiles of convolved models using {\PSFVa} and {\PSFIa} ({\PSFVb} and \PSFIb), the purple line used \PSFK. Three different symbols and error bars show measured values: $\bullet$ $R$ band (\citealt{SaMoHaBo:94}; \MBH), $\star$ $V$ band (\Lequeux), and from profiles on both sides of the disc $\circ$ $6660${\AA} band (\Zheng). The $R$-band model was convolved with the measured {\PSFMBH} (including lower and upper errors) to produce the white line (cyan-coloured region). The lower limiting radius $r_{110}$ -- where the convolved models using {\PSFVa}, {\PSFIa}, and {\PSFK} lie $\ge\!10$ per cent above the input model -- is marked with a coloured bullet with a black border. \textbf{b}) Three colour profiles $R-i$ are shown for: the model (thick solid line), the convolved model using {\PSFVa} and {\PSFIa} (solid line), and the convolved model using {\PSFVb} and {\PSFIb} (dash-dotted line).}
\label{fig:ngc5907}
\end{figure*}

\subsection{Models of the surface-brightness structure}\label{sec:ngc5907model}
I modelled NGC 5907 with $h_{\text{r}}=90\arcsec$, $z_{0}=15\arcsec$, $n_{\text{s}}=2$, and assumed $D=11.7\,\mbox{Mpc}$. In comparison, {\MBH} use the values $h_{\text{r}}=90\arcsec$ and $z_{0}=16\arcsec$ (which corresponds to $h_{\text{z}}=z_{0}/2=8\farcs1$; they specify all parameters in kpc and assume $D=11\,\mbox{Mpc}$). Model and observed vertical-axis profiles are shown in Fig.~\ref{fig:ngc5907}a, and corresponding $R-i$ colour profiles in Fig.~\ref{fig:ngc5907}b. The five PSFs that are used in the $R$ band cause intensity (colour) differences of up to $1.5$ ($0.3$) {\magg} at $r=70\arcsec$. At the same radius the surface-brightness (colour) profiles of the convolved models are 2.2--3.7 (0.96--1.5) {\magg} brighter than the input model value. The $R$-band \citep{SaMoHaBo:94}, $6660${\AA}-band (\Zheng), and $V$-band (\Lequeux) measurements fall between the convolved profiles of {\PSFVa} and {\PSFVb} throughout most of the radial range. $V$-band and $R$-band data nearly overlap where $r\la30\arcsec$, and differ by up to about $1\,\magg$ at larger radii in the halo.

For $30\la r\la95\arcsec$, the halo that {\MBH} measure is up to about $1\,\magg$ brighter than the model that was convolved with {\PSFMBH}. According to the authors, the halo is not scattered light. However, the conclusion is different with {\PSFMBH} using its upper error bars, and with {\PSFMBHT}, which convolved models nearly overlap all measured values, but the two ones at $r=74\arcsec$ and $r=83\arcsec$. The upturn in the outer part of {\PSFMBH} and {\PSFMBHT} causes a shelf of nearly constant brightness in the convolved model profile where $80\la r\la115\arcsec$ -- there is no corresponding shelf in the measurements\footnote{It was necessary to extrapolate {\PSFMBH} and {\PSFMBHT} with a decreasing slope outside their largest radii to avoid positive slopes in the outer regions of convolved model profiles.}. Considering the difficulties in measuring {\PSFMBHT} accurately at both smaller and larger radii (Sect.~\ref{sec:arngc5907}), it cannot be excluded that all of the measurements can be explained as scattered light. {\PSFMBHT} is particularly noisy in the range $70\la r\la95\arcsec$, which delimits the accuracy of two values that are brighter than the convolved model. Under the assumption that the measurements are scattered light, at $z=70\arcsec$ the required S/N of the measurements to measure the 2.5\,{\magg} fainter input model is 10. The accuracy requirement quickly grows to enormous values at larger distances, where it is impossible to measure the exponential structure (also see Sect.~\ref{sec:toy}).

According to \Zheng, the $0.3\,\magg$ offset, which they adopt between the $R$-band and the $6660${\AA}-band measurements, occurs due to different filter bandwidths. Under the assumption that the halo is scattered light, varying PSFs justify deviating measurements, also the $1\,\magg$ fainter $V$-band measurements of \Lequeux. I have no $J$-band or $K$-band PSFs to model the near-infrared observations of \citet{RuWoHo.:97} or \citet{JaCa:98}, but the PSFs of these bands -- as measured with, for example, data of the 2MASS Large Galaxies Atlas \citep[][where observations were made at Mt.\ Hopkins in Arizona and the Cerro Tololo Inter-American Observatory]{JaChCu.:03} -- are even brighter at larger radii than the $I$-band PSF \citep{Mi:07}, and scattered-light effects are therefore even stronger.

The $R-i$ colour profiles of the convolved models in Fig.~\ref{fig:ngc5907}b show red excess where $z>18\arcsec$, in qualitative agreement with the $V-I$ profile of {\Lequeux} (see their fig.~3). The difference between the colour profiles that were derived using {\PSFVa} and {\PSFIa}, versus {\PSFVb} and {\PSFIb}, illustrate that derived colours in the halo, which is dominated by scattered light, completely depend on the PSFs in the used bandpasses. Whilst {\PSFVa} and {\PSFIa} contain fainter wings than {\PSFVb} and {\PSFIb}, their radial difference is greater, which result in more red excess in the halo.

\section{Discussion and conclusions}\label{sec:conclusions}
I have presented a first detailed overview and comparison of already measured radially extended PSFs, since the first summary of {\King} and the more recent summary of \Bernstein. The overview demonstrates that all PSFs are expected to continue to large radii (angles), and that our knowledge is poor of how the PSF varies with the time and the wavelength, in particular at large radii. \citet{Va:58} and later {\King} (amongst a few others), find that the PSF declines as an $r^{-2}$ power-law at large distances. {\Bernstein} presents a summary of PSFs to show that various measurements contradict the use of a general $r^{-2}$ slope. I have normalized all PSFs and put them in the same plot, and the result shows no evidence against an $r^{-2}$ power-law slope at large radii. Instead, several radially extended measurements point at a more shallow decline. There is one exception, the PSF of \citet{AbDo:14}, which is measured with the only configuration that uses refractive instead of reflective optics. Encircled energy plots show significant amounts of light in the far wings of the PSF, and there is currently no clear correspondence between the amount of encircled energy in the outer parts of the PSF and atmospheric extinction.

Early studies measured the outer parts of the PSF with observations of planets, the Moon, and the Sun, instead of using (only) stars. I have shown that extended objects can be regarded as point sources, beginning at a geometry-dependent radius. Complementary observations of these comparatively bright objects will make possible more accurate measurements of varying and radially extended PSFs. An additional point of concern that I have not addressed here, is how the PSF changes across the surface, away from the optical axis (\Slater); new PSF measurements need to take such issues into account as well.

I have studied the role of scattered light to shape the surface-brightness structure of three models of edge-on disc galaxies; the model set consisted of a small galaxy, an intermediate-sized galaxy, and a large galaxy, which all contain a single thin disc. I have used {\PSFK} (\King) as a lower limit, and {\PSFVa} \citep{Mi:02} as a slightly-above average indicator of scattered-light effects. The results show that in the case of edge-on galaxies it is necessary to use a PSF that extends to at least 1.5 times the measured radius, regardless of the size of the galaxy, or the measurements cannot be corrected; it seems that it would be safe, for any measurement, to use a PSF of twice the measured radius. It is notable that the requirements on the accuracy of the measurements at large vertical distances quickly becomes insurmountable to measure an underlying exponential component.

Surface-brightness profiles of edge-on galaxies are divided into four components: a centre bulge, a thin disc, a thick disc, and a halo. The first faint diffuse halo around an edge-on disc galaxy was found around the Sc-type NGC 5907. Examples of explanations to the origin of the halo include that it traces the dark-matter halo \citep{SaMoHaBo:94}, and that parts or all matter in the halo could be low-mass stars (\citealt{SaMoHaBo:94}; \Lequeux). \citet{ZeLiMa.:00} find one halo star in their \textit{HST} $H$-band observations of NGC 5907, instead of the expected 100; they favour a stellar population with a very high dwarf-to-giant ratio to explain the halo. \citet{YoBoKa.:00} present near-infrared observations from above the Earth atmosphere, and rule out hydrogen-burning stars as a possible cause of the halo. \citet{ShZhBr.:98}, {\Zheng}, and \citet{MaPeGa.:08} discover and discuss comparatively faint stellar tidal streams that extend far out from NGC 5907 on its north-east and south-west sides. The influence of the tidal streams on the results presented here is minor, despite that the streams partly overlap the regions that are used to measure the galaxy profile and the PSFs; the reason is that the relative intensity between the streams and the background is very small, and even smaller than the errors in the reduced image of {\MBH}. {\Zheng} also find that vertical surface-brightness structures away from the minor axis are asymmetric, and conclude that the halo is not real, but is contaminated by light from the stream and residual light from field stars. They do not consider scattered light from the galaxy itself.

Scattered light was dismissed as an explanation to the halo by {\MBH} after a careful analysis of the PSF, which later studies of this galaxy do not address with as much care. I have reassessed the original analysis of NGC 5907, regarding the role of scattered light, and I have come to the conclusion that it likely is the major reason to the appearance of the halo. Specifically, I have analysed the influence of the variations in the galaxy background. I have lowered the background value by 3\,ADU (0.17\%), compared to the value that was seemingly used to measure the extended original {\PSFMBH}. I have then measured a significantly brighter {\PSFMBHT}, where the new PSF is still mostly within the error bars of \PSFMBH. {\MBH} quote the accuracy of their sky data as $10.3\,\text{ADU}$, which is more than three times as large as the background offset that I applied. The new finding illustrates the high accuracy that is required of both the PSF and the measurements when dealing with scattered light.

Finally, the $R-i$ colour profiles of the scattered-light dominated halo of the models show very strong red excess, which is all caused by the shape of the PSFs. Notably, {\Jong} comes to the same conclusion, based on models that are convolved with less extended PSFs. It seems that the CCD-specific red-halo effect \citep{SiClHa.:98} plays a strong role to enhance the red excess in the $i$-band.

I emphasize that the object asymmetry and the varying background complicate the modelling. Measurements of other authors of NGC 5907 agree with this conclusion. In particular, it appears that the observations of {\Lequeux} were made with a fainter $V$-band PSF that is similar to \PSFK, whilst {\PSFMBHT} is brighter than {\PSFVa} throughout most of the radial range. Assuming that the halo is induced by the PSF, there is also no argument against a larger offset value between the $R$ and the 6660{\AA} bandpasses that {\Zheng} use. In view of the alternative explanations to the bright measurements in the halo of NGC 5907, the one of scattered light is simple, and does not require any exotic stellar populations. Despite very small adjustments to values that were used in the analysis, the impact on physical results is large.

With his study on two sets of smaller edge-on galaxies, {\Jong} points out that effects of scattered light in observations of edge-on disc galaxies can be significant. The conclusion of this paper is that these effects can be even stronger -- faint regions around edge-on galaxies of all sizes are affected by scattered light. I have indicated that PSFs generally vary with time and wavelength, and that also their faint outer wings contribute significantly to observed structures. The question is, in view of this new awareness, how observations of astronomical objects and their faint structures in general are affected by scattered light? In a second paper, I will study the influence of scattered light on a larger set of models and observations of different types of galaxies.

\begin{acknowledgements}
I am most grateful to H.\ Morrison, who provided me with the originally reduced images of NGC 5907. I thank R.\ Abraham for sending me the Dragonfly Array PSF data. I thank A.\ Partl, A.\ Monreal-Ibero, R.\ de Jong, C.\ Vocks, and M.~M.\ Roth for comments on the manuscript. This research was supported by funds of PTDESY-05A12BA1.
\end{acknowledgements}

\bibliographystyle{aa}

\appendix

\section{About \textit{HST} PSFs}\label{sec:HSTpsf}
PSFs that are measured with the \textit{HST} -- and other space-based telescopes -- are less affected by dust and the Earth atmosphere than PSFs of ground-based telescopes, and they are often much more compact. The tool \textsc{Tiny Tim} \citep{KrHoSt:11} was created to provide model PSFs for all instruments and most observing modes of the \textit{HST}. The tool, its documentation, and additional information are available at the project web site\footnote{The \textsc{Tiny Tim} web site at the STScI: \href{http://tinytim.stsci.edu}{http://tinytim.stsci.edu}.}. \textsc{Tiny Tim} makes a best effort to account for all factors in the optical path that affect the PSF (see the \textsc{Tiny Tim} user guide). The resulting PSFs are many times of good quality, considering that many of these factors vary with time or field position. One comment below the `Performance' section at the project web site is relevant to this study:
\begin{quotation}
For PSF subtraction where details of the outer portions of the distribution are important, and for other cases where accurate PSF characterizations are needed, these modelled PSFs may lack suitable accuracy, and empirical PSF techniques should be considered when practical.
\end{quotation}
This is a highly important comment, considering that in this paper I demonstrate the decisive influence of the outer faint regions of the PSF to induce haloes and affect colour profiles.

I mention additional content of two sections in the \textsc{Tiny Tim} user guide (version 6.3) that are related to this issue. Light is scattered into the outer regions of the PSF of the second Wide Field and Planetary Camera \citep[WFPC2;][]{KrBu:92,Kr:95}; outer regions of the PSF will be underestimated since this effect is not included in \textsc{Tiny Tim}. Furthermore, due to a defect in the high-resolution CCD (HRC) of the Advanced Camera for Surveys (ACS), a halo is created that surrounds the PSF for wavelengths $\lambda>600$nm; \textsc{Tiny Tim} provides a first estimate of the halo effect. In the ACS instrument handbook\footnote{The ACS instrument handbook is available at:\\ \href{http://www.stsci.edu/hst/acs/documents/handbooks/current/cover.html}{http://www.stsci.edu/hst/acs/documents/handbooks/current/cover.html}.}, the halo is said to appear for $\lambda>700$nm (HRC) and $\lambda>900$nm [wide field camera (WFC)].

\citet{SiJeBe.:05} present measurements of the ACS PSFs where they consider $r=5\farcs5$ a safe extent to measure all flux within an `infinite' aperture; they treat all filters of both the HRC and the WFC. I show in Sect.~\ref{sec:toy} how important it is to use a PSF that is not truncated or underestimated at large radii, and I advocate the use of a PSF that is at least 1.5 times as extended as the vertical distance of the edge-on galaxy. It would be valuable to see how an extension of the PSF to, say, $r=10\arcsec$ would affect integrated scattered light in a study that is similar to what I present here for ground-based telescopes -- this would show if radially extended PSFs make any difference also with \textit{HST} data.

\section{Details of the extrapolation and normalization of the discussed PSFs}\label{sec:aenpsf}
I extrapolated {\PSFK}, {\PSFVa}, and {\PSFVb} to $r=90^{\circ}$, assuming a power law dependence $r^{-2}$; in agreement with the existing outer parts of \PSFK, \PSFPi ($\blacktriangle$), and \PSFS. Studies that report on contradicting slopes present a noisy outer PSF that does not extend very far (\PSFB, SDSS PSFs). 

The spectroscopic {\PSFPa} (seeing $0\farcs8$) does not extend far; it was extrapolated with {\PSFC}, as the two PSFs match well in the range $4\le r\le25\arcsec$. I could shift three PSFs precisely, with available information on how they overlap other PSFs: {\PSFKo}, {\PSFPi} (which triangle-symbol points overlap the aureole line of {\PSFK}), and {\PSFVa}. \PSFVb, \PSFIa, and {\PSFIb} are fixed relative to \PSFVa. {\PSFS} is not measured at small radii, which is why I offset it by $+0.25\,\magg$, to match it with the other PSFs, the true value could be slightly lower or higher. \citet{AbDo:14} fix the offset between {\PSFAbr} and {\PSFS}, I used the same value. I offset the SDSS PSFs by hand, to have them agree with the integrated PSFs. These PSFs are poorly determined near their outer limit at $r\simeq30$--$60\arcsec$, which makes an accurate extrapolation to larger radii difficult. {\PSFMBH} is also poorly determined at large radii; I offset it by $-13.6\,\magg$ to have this large-seeing PSF slightly fainter at the centre than {\PSFVa}. I used the same offset with {\PSFMBHT}.

The measurements of PSF$_{\text{S74}}$ and PSF$_{\text{M92}}$ were offset by $-7\,\magg$ to overlap \PSFK, and all three versions of PSF$_{\text{G05}}$ were offset by $-12\,\magg$ to overlap \PSFIa. These three PSFs were not measured in the centre, and the true offsets could be both slightly larger and slightly smaller. Finally, the (centred) {\PSFB} was offset to agree with {\PSFK} in the centre region.

\section{Details of the analysis of NGC 5907}\label{sec:dngc5907}
This section contains supplementary details to the discussion of the original analysis and the reconsidered analysis of NGC 5907 in Sect.~\ref{sec:ngc5907}.

\begin{figure*}
\sidecaption
\includegraphics{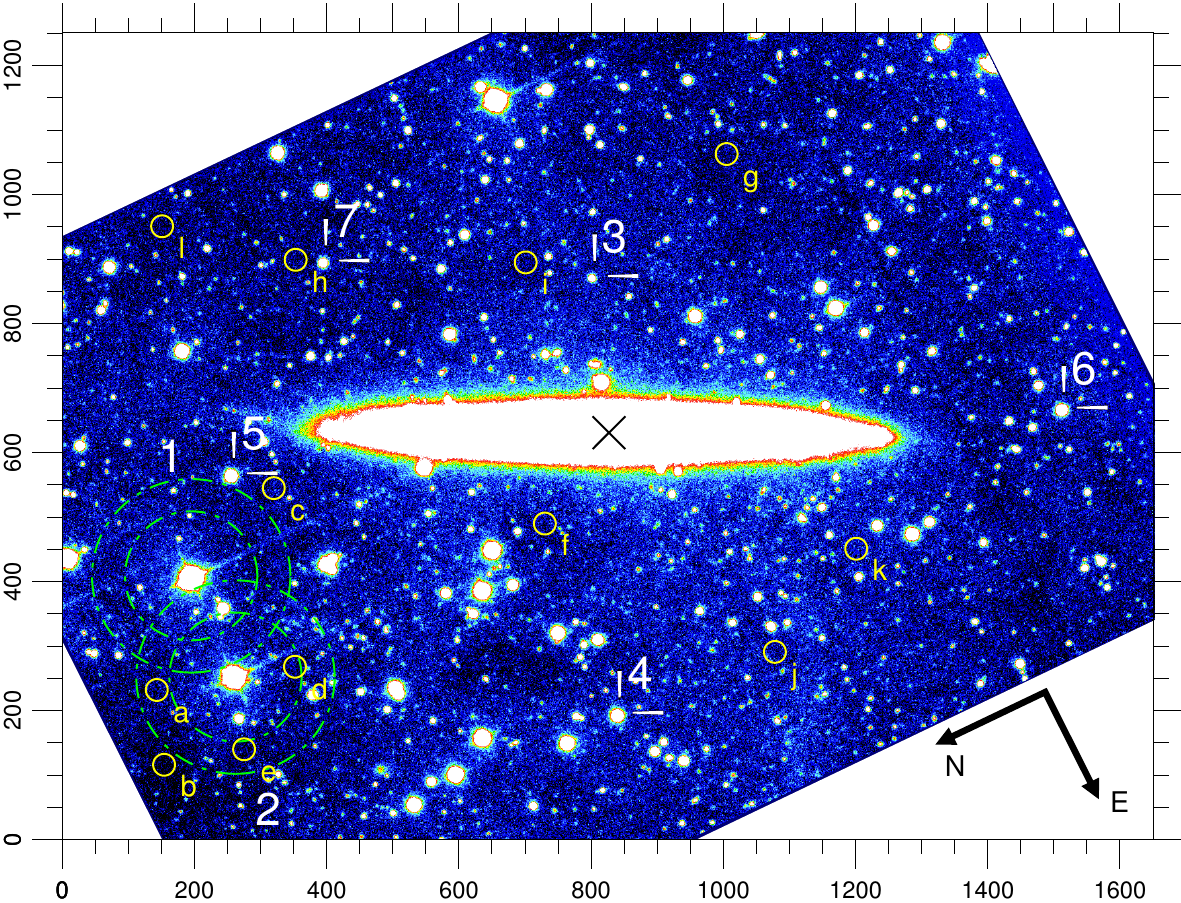}
\caption{The originally reduced image of {\MBH} for NGC 5907, shown with a colour map that emphasizes the variations of the background field (their fig.~1b). The unit of the axes is pixels. The used galaxy centre position is marked with an $\times$. Two sets of green circles indicate the distance at 100 and 150\,px away from the two saturated stars 1 and 2. The positions of five bright stars (twelve regions that were used to measure the background) are indicated with markers and a number (circles and a character), cf.\ Table~\ref{tab:brightPSF} (Table~\ref{tab:backgrounds}).}
\label{fig:ngc5907im}
\end{figure*}

\begin{figure*}
\centering
\includegraphics{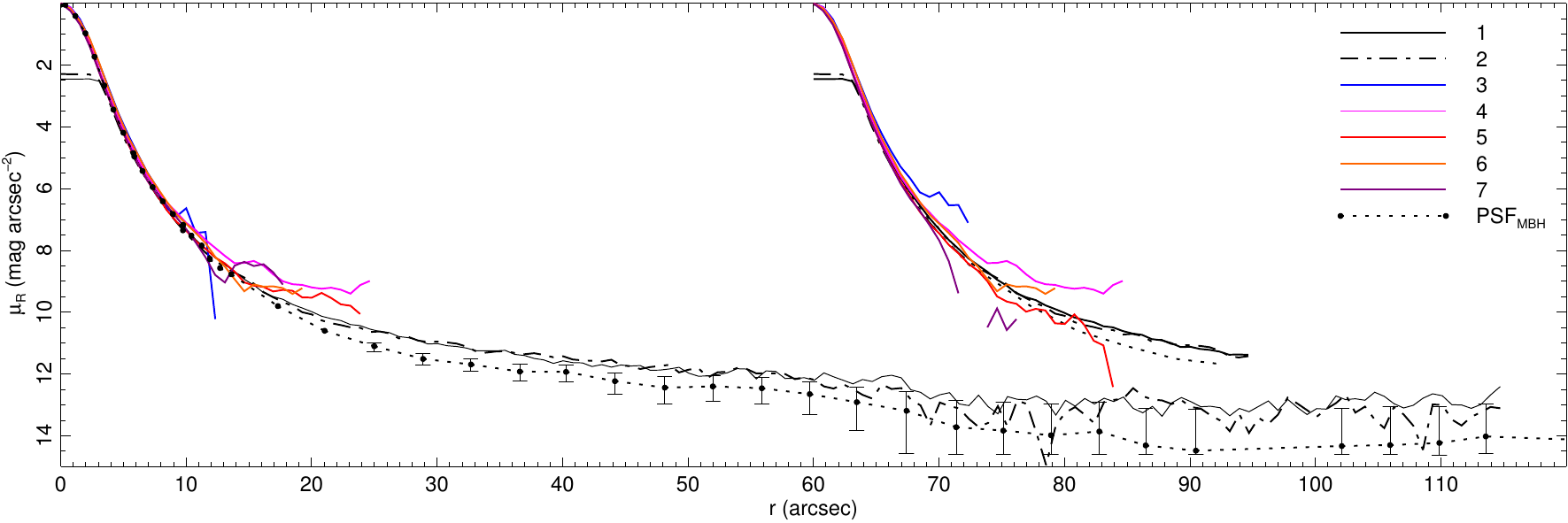}
\caption{PSFs that were derived from the reduced $R$-band image of NGC 5907 (Fig.~\ref{fig:ngc5907im}). The azimuthal average is drawn as magnitude versus radius. The left-hand set of lines used individually set background values. The set of lines that are offset by 60{\arcsec} were all derived using the background value 1760\,ADU. {\PSFMBH} is drawn with a dotted line, together with its error bars. The saturated-star PSFs are drawn with black solid (1) and dash dotted (2) lines -- {\PSFMBHT} is the average of these two PSFs. The PSF of the remaining five bright stars are drawn with coloured lines out to a maximum radius $r_{\text{max}}$, cf.\ the legend and Table~\ref{tab:brightPSF}.}
\label{fig:ngc5907npsf}
\end{figure*}

\subsection{Details of the original observations}\label{sec:aongc5907}
To make possible a careful flat fielding of their data, {\MBH} took 22 1800\,s exposures of the sky during the five nights of the observing run; these sky regions were offset some $1\degr$ from the galaxies of their study. Out of the 22 frames they use 15 frames that are judged to be free from bright stars or defects that could compromise the flat field. The sky frames have about 1760\,ADU per pixel, which ought to produce a flatfield where pixel-to-pixel errors are smaller than about 0.29 per cent (5.1\,ADU). They make a more detailed error model of the sky, and derive an error of the rotated and shifted sky-image pixels as 0.60 per cent ($10.5\,\text{ADU}$), which agrees well with their empirically measured value 0.59 per cent ($10.3\,\text{ADU}$).

{\MBH} use several bright, but not saturated, stars, to measure the inner parts of a PSF. They measure the outer PSF with the two bright and saturated stars that are seen to the lower left of the galaxy disc in their fig.~1; their figure is reproduced here in Fig.~\ref{fig:ngc5907im}. To this end they prepare and apply a mask that removes all stars, diffraction spikes, and background galaxies in the field around each star. The unmasked pixels are averaged in radial annuli around the centre, starting at $r=20\,\text{px}$. The resulting {\PSFMBH} is reproduced in Fig.~\ref{fig:psf} (and also with its error bars in Fig.~\ref{fig:ngc5907npsf}), it extends out to $r=116\arcsec$.

Later, {\Lequeux} claim to measure a PSF that reaches $10^{-7}$ of the peak intensity at $r=16\arcsec$ (the PSF is not published). This PSF is likely incorrectly measured, as it is \citep[along with the SDSS PSFs of][]{ZiWhBr:04} much steeper than any of the other PSFs in Figs.~\ref{fig:psf} and \ref{fig:bpsf}. {\Zheng} also claim to derive a PSF, but there is no information about how far radially and deeply it extends.

The vertical surface-brightness structure of NGC 5907 is measured after dust and stars are masked, and remaining pixels are averaged in 100 pixel wide strips, which heights change exponentially from 5\,px near the major axis to a maximum of 99\,px at the vertical distance $z=160\arcsec$. The mask they use is shown in fig.~3 in \MBH.

\begin{figure*}
\includegraphics{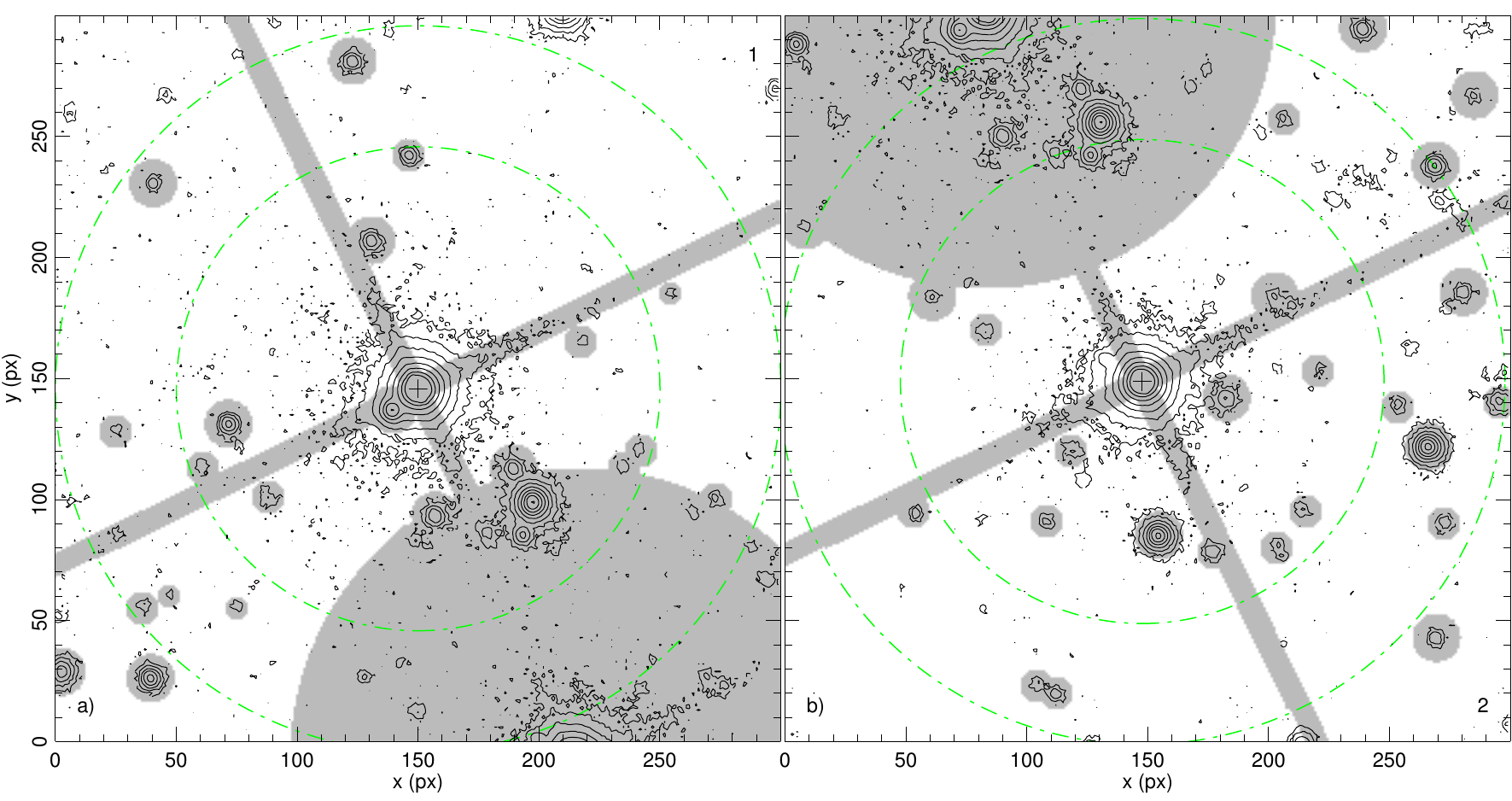}
\caption{Contour plots of the two saturated stars \textbf{a}) 1 and \textbf{b}) 2 in Fig.~\ref{fig:ngc5907im}, where masked regions are marked in grey. The grey tilted crosses indicate the masked diffraction spikes. The green circles mark the radii 100 and 150\,px.}
\label{fig:ngc5907satstars}
\end{figure*}

\begin{table}
\caption{Image statistics of selected background regions in Fig.~\ref{fig:ngc5907im}.}
\centering
\label{tab:backgrounds}
\tabcolsep4.1pt
\begin{tabular}{l@{\ \!}rrlllllc}\hline\hline\\[-1.8ex]
Id & \multicolumn{1}{c}{$x$} & \multicolumn{1}{c}{$y$} & $\min(z)$ & $\max(z)$ & \multicolumn{1}{c}{$\overline{z}$} & $\text{med}(z)$ & \multicolumn{1}{c}{$\sigma_z$} & \multicolumn{1}{c}{T}\\\hline\\[-1.8ex]
\textit{a} &  143 &  232 & 1737.0 & 1783.3 & 1759.1 & 1759.3 & 7.1 & x\\
\textit{b} &  154 &  116 & 1728.8 & 1778.5 & 1753.5 & 1753.2 & 6.5 & x\\
\textit{c} &  320 &  545 & 1737.2 & 1787.5 & 1760.4 & 1760.4 & 7.0 & $\circ$\\
\textit{d} &  352 &  268 & 1732.1 & 1806.9 & 1759.7 & 1759.3 & 7.8 & $\bullet$\\
\textit{e} &  275 &  140 & 1736.7 & 1777.1 & 1757.9 & 1757.9 & 6.8 & x\\
\textit{f} &  730 &  490 & 1736.8 & 1783.2 & 1761.8 & 1761.9 & 6.9 & -\\
\textit{g} & 1005 & 1063 & 1734.3 & 1785.7 & 1759.5 & 1759.4 & 6.7 & -\\
\textit{h} &  353 &  899 & 1733.8 & 1780.0 & 1758.5 & 1758.6 & 6.7 & $\circ$\\
\textit{i} &  701 &  895 & 1739.6 & 1783.3 & 1759.9 & 1759.8 & 6.5 & $\circ$\\
\textit{j} & 1078 &  291 & 1734.7 & 1790.6 & 1762.3 & 1762.1 & 7.0 & -\\
\textit{k} & 1201 &  451 & 1739.8 & 1790.6 & 1764.2 & 1764.1 & 6.9 & -\\
\textit{l} &  151 &  951 & 1732.2 & 1782.4 & 1757.6 & 1757.6 & 6.7 & -\\[0.3ex]
\hline
\end{tabular}
\tablefoot{Column~1, identifier; Cols.~2 and 3, $x$- and $y$-positions of ring centre (px.); Cols.~4--8, minimum, maximum, mean, median, and standard-deviation values (ADU) of, in each case, 1257 pixels $z_{i}$ inside a ring with the radius 20\,px that is centred on the coordinates ($x,y$); and Col.~9, four symbols specify if the region falls on top of the (T)idal streams around NGC 5907 that are seen in fig.~2 in \citet{MaPeGa.:08}: x outside of the field, - away, $\circ$ partly on top, $\bullet$ right on top.}
\end{table}

\begin{table}
\caption{Bright stars in Fig.~\ref{fig:ngc5907im} that were used to derive a PSF.}
\centering
\label{tab:brightPSF}
\begin{tabular}{lrrlrlc}\hline\hline\\[-1.8ex]
Id & \multicolumn{1}{c}{$x$} & \multicolumn{1}{c}{$y$} & \multicolumn{1}{c}{bg.} & \multicolumn{1}{c}{$r_{\text{max}}$} & colour & T\\\hline\\[-1.8ex]
1 &  195 &  409 & 1759   & 115.5 & black solid & $\circ$\\
2 &  262 &  252 & 1759   & 115.5 & black dash dotted & $\circ$\\
3 &  801 &  870 & 1762   & 13 & blue & $\circ$\\
4 &  840 &  193 & 1760   & 25 & magenta & -\\
5 &  256 &  565 & 1758   & 24 & red & $\bullet$\\
6 & 1164 & 1126 & 1760   & 20 & orange & x\\
7 &  395 &  894 & 1758   & 18 & purple & $\bullet$\\[0.3ex]
\hline
\end{tabular}
\tablefoot{Column~1, identifier; Cols.~2 and 3, $x$- and $y$-positions in the reduced image (px.); Col.~4, used background value (ADU); Cols.~5 and 6, maximum radius (arcsec) and line colour in Fig.~\ref{fig:ngc5907npsf}; and Col.~7, four symbols specify if the region falls on top of the (T)idal streams around NGC 5907 that are seen in fig.~2 in \citet{MaPeGa.:08}: x outside of the field, - away, $\circ$ partly on top, $\bullet$ right on top.}
\end{table}

\subsection{Details of the reconsidered analysis}\label{sec:arngc5907}
Following the approach of {\MBH}, I derived a PSF by averaging unmasked pixels in radial annuli around the centre of bright stars. Whilst {\MBH} quote a background value of 1760\,ADU in the sky image, that value does not seem to apply in all parts of the galaxy image. To quantify the differences I measured background values in different regions in Fig.~\ref{fig:ngc5907im}; image statistics of twelve regions \textit{a--l} are shown in Table~\ref{tab:backgrounds}. In the same table I also marked whether the respective region falls on top of the tidal streams around NGC 5907, as are visible in \citet{MaPeGa.:08}.

This test shows slightly lower values on the left-hand side of the image, and near the two saturated stars. As a compromise, I estimated that the average background of the regions \textit{a}, \textit{d}, and \textit{e} near the lower saturated star (2) is 1759\,ADU. This value is a bit uncertain, and it is possible that it varies by, say, about 2\,ADU in the radial range away from each saturated star. The same values also illustrate the difficulties in measuring a PSF at the level of the background with these data, in particular for radii $r\ga100\,\text{px}$ where all three background regions fall inside the measured region; here, scattered light from star 2 makes these values higher than the sky background. The two stars are also situated right next to a tidal stream around NGC 5907 \citep[compare the position of region \textit{d} with fig.~2 in][]{MaPeGa.:08}, but this is not obvious when comparing the values of the separate regions.

The background is brighter on the right-hand side of the image, where I measured $1759.5$--$1764.2\,$ADU in the regions \textit{g}, \textit{i}, \textit{j}, and \textit{k}. None of these regions falls directly on top of a tidal stream. For comparison, in the other parts of the galaxy, {\MBH} measure a lower value in the region that is offset by 100\,px (4.1\,kpc) to the left of the galaxy centre, which they attribute a flat-fielding defect. In this region, \textit{f}, I measured the mean value 1761.8\,ADU, which is slightly lower than in the regions \textit{j} and \textit{k}, but still significantly higher than the mean value 1759\,ADU around the lower saturated star 2. All background values that are quoted here fall within the statistical accuracy of 5--10\,ADU.

I derived a PSF for five bright stars in Fig.~\ref{fig:ngc5907im}, using both 1760\,ADU and an individually selected value as background, cf.\ Table~\ref{tab:brightPSF}; the PSF of each of these relatively faint stars is only useful out to an approximative radius that I refer to as $r_{\text{max}}$. I also derived more extended PSFs for the same saturated stars as \MBH. The brighter contour levels and the regions that I masked around these two stars are shown in Fig.~\ref{fig:ngc5907satstars}. The original {\PSFMBH} is shown together with the new PSFs of all seven stars in Fig.~\ref{fig:ngc5907npsf}.

PSFs that are measured in different parts of the image should overlap (neglecting spatial differences across the field such as reported by \Slater, and which magnitudes are in any case unknown here). The PSFs of stars 1 and 2 are slightly closer to {\PSFMBH} when the background value 1760\,ADU is used, and they nearly overlap when the value is instead 1762\,ADU (not shown). The PSF of star 3, which is located above the galaxy disc in Fig.~\ref{fig:ngc5907im}, lies above the other PSFs in the region $r\ga5\arcsec$. It overlaps the other PSFs better when the background is set to 1762\,ADU, which seems to be an appropriate background value in the neighbourhood of this star. The stars 4 and 6 are located right and below of the galaxy disc; their PSFs seem to overlap the other PSFs using the background value 1760\,ADU. The stars 5 and 7 are located to the left side of the galaxy disc. The background level of both these stars seems to be closer to 1758\,ADU, as in this case they overlap the PSFs of stars 1 and 2. The background values are low, even though both stars lie directly on top of a tidal stream. Both PSFs are markedly fainter if the value 1760\,ADU is used instead.

The PSFs of stars 1 and 2 overlap each other closely for $r<70\arcsec$. For larger radii the surface-brightness slope is positive; the overestimated measurements are there influenced by the PSF wings of other field stars as well as the varying background. The same conclusion applies as a plausible cause to the upwards slope in the outermost parts of \PSFMBH. The two PSFs are about $0.5\,\magg$ brighter than {\PSFMBH} at both $r=40\arcsec$ and $r=60\arcsec$. Their average, {\PSFMBHT}, is poorly determined for $r>70\arcsec$. It is worth noting that the fainter parts of the surface-brightness structure of NGC 5907 depend on these far regions of the PSF.

The entire difference between {\PSFMBH} and {\PSFMBHT} occurs due to a background level that is 3--4\,ADU (0.17--0.23 per cent) lower than the value that {\MBH} appear to use (1762--1763\,ADU). This test indicates the high accuracy that is required in the background to derive a PSF with the two used relatively faint stars; it is higher than the expected accuracy of the flatfield image they use, 5--10\,ADU (0.29--0.59 per cent). The data and the resulting {\PSFMBHT} are not accurate enough to deconvolve the galaxy measurements accurately. The reconsidered {\PSFMBHT} is also drawn in Fig.~\ref{fig:psf}.

To test the influence of masking or not masking the diffraction spikes, I calculated an additional PSF where these were masked (the affected regions are indicated by the tilted cross in Fig.~\ref{fig:ngc5907satstars}). This test shows that the difference due to the diffraction spikes is small. The masked PSF is about $0.1\,\magg$ fainter than the unmasked PSF at $r=22\arcsec$, which decreases to near zero at $r=40\arcsec$ (not shown).

For completeness, I calculated a vertical surface-brightness profile below (East-North-East of) the galaxy disc that is nearly identical to that of {\MBH}; to this effect I set the galaxy centre to $(x,y)=(827,631)\,\text{px}$ (this position is indicated in Fig.~\ref{fig:ngc5907im}), the background value to 1760\,ADU, and used the same mask. A slightly different profile results if the galaxy centre is instead set to the peak of the surface-brightness profile at $(x,y)=(827,628)\,\text{px}$. The galaxy surface-brightness structure is, moreover, not perfectly symmetric due to (horizontal) dust lanes, which are present in particular above the galaxy disc (these cannot be seen in Fig.~\ref{fig:ngc5907im}). Tidal streams above the galaxy \citep[see fig.~2 in][]{MaPeGa.:08} plausibly contribute to the asymmetry as well. The vertical profile that is measured above the disc, furthermore, differs slightly from the profile that is measured below the disc. Also, the used mask is unable to account for extended wings of the PSFs of numerous field stars around the galaxy and near the two saturated stars. However, as in the case of the two saturated stars, the biggest uncertainty to the (outer parts of the) galaxy profile is also here the background value, and the other effects are secondary to this effect. Therefore, in the discussion in Sect.~\ref{sec:ngc5907model} I used the profile of {\MBH}.

\listofobjects

\end{document}